\documentclass[12pt]{article}
\usepackage{epsf}
\usepackage{color}
\usepackage{graphicx}
\usepackage{amsmath}
\usepackage{amssymb}
\usepackage{latexsym}

\begin{document}
\newcommand{\mathcalpl}{M_{\mathcalathrm{Pl}}}
\setlength{\baselineskip}{18pt}

\begin{titlepage}
\begin{flushright}
OU-HET 694/2011 
\end{flushright}

\vspace*{1.2cm}

\begin{center}
{\Large\bf Dirichlet Higgs as radion stabilizer\\in warped compactification
}
\end{center}

\lineskip .75em
\vskip 1.5cm

\begin{center}
{\large 
Naoyuki Haba$^{a,}$\footnote[1]{E-mail:
\tt haba@phys.sci.osaka-u.ac.jp}, Kin-ya Oda$^{a,}$\footnote[2]{E-mail:
\tt odakin@phys.sci.osaka-u.ac.jp}, 
and Ryo Takahashi$^{b,}$\footnote[3]{E-mail: 
\tt ryo.takahashi@mpi-hd.mpg.de}
}\\
\vspace{1cm}

$^a${\it Department of Physics, Graduate School of Science, Osaka
University, \\
Toyonaka, Osaka 560-0043, Japan}\\

$^b${\it Max-Planck-Institut f$\ddot{u}$r Kernphysik, Saupfercheckweg 1, \\ 
69117 Heidelberg, Germany}\\

\vspace*{10mm}

{\bf Abstract}\\[5mm]
{\parbox{13cm}{\hspace{5mm}
We study implications of generalized non-zero Dirichlet boundary condition 
along with the ordinary Neumann one on a bulk scalar in the Randall-Sundrum 
warped compactification. First we show profiles of vacuum expectation value of 
the scalar under the general boundary conditions. We also investigate 
Goldberger-Wise mechanism in several setups with the general boundary 
conditions of the bulk scalar field and find that the mechanism can work under 
non-zero Dirichlet boundary conditions with appropriate vacuum expectation 
values. Especially, we show that $SU(2)_R$ triplet Higgs in the bulk 
left-right symmetric model with custodial symmetry can be identified with the 
Goldberger-Wise scalar.
}}
\end{center}

\end{titlepage}

\section{Introduction}
One of the main targets of the CERN Large Hadron Collider (LHC) is the 
discovery of an evidence of extra-dimension(s) as well as the Higgs particle.
There are extra dimensional alternatives to the ordinary electroweak symmetry 
breaking (EWSB) mechanism in the Standard Model (SM), such as the gauge-Higgs
 unification (GHU) 
\cite{Fairlie:1979at,Fairlie:1979zy,Manton:1979kb,Hosotani:1983xw}, the little 
Higgs \cite{ArkaniHamed:2001nc}, the Higgsless \cite{Csaki:2003dt}, and the 
Dirichlet Higgs \cite{Haba:2009pb} models\footnote{See also Refs. 
\cite{Serone:2005ds,Serone:2009kf}, \cite{Schmaltz:2005ky,Perelstein:2005ka}, 
\cite{Csaki:2005vy,Simmons:2006iw}, and references therein for nice reviews of 
the GHU, little Higgs, and Higgsless models, respectively.}, and so on. 
Possible approaches to address the Higgs mass hierarchy problem are large 
extra-dimension scenario \cite{ArkaniHamed:1998rs,Antoniadis:1998ig} and the 
Randall-Sundrum (RS) model \cite{Randall:1999ee} in addition to supersymmetry.

The extra-dimensional models can also give phenomenologically interesting 
features and predictions,\footnote{First proposal of TeV scale compactification
 is made in \cite{Antoniadis:1990ew}.} for example, a candidate for dark matter
 (DM) from the Universal Extra-Dimensions (UED) model \cite{Appelquist:2000nn} 
and deviations of couplings of the Higgs in the context of GHU scenario 
\cite{Hosotani:2008tx,Hosotani:2008by}, brane localized Higgs potential models 
\cite{Haba:2009uu,HOT3}, and Dirichlet Higgs model 
\cite{Haba:2009pb,Haba:2010xz,Oda:2010te}\footnote{An implementation of such 
deviations to flavor physics has been discussed in \cite{Holthausen:2009qj}, 
which can also give a DM candidate.}. Constructing realistic models in the 
warped five-dimensional spacetime proposed by Randall and Sundrum is still an 
interesting issue. After this proposal, Goldberger and Wise (GW) presented a 
mechanism for stabilizing the size of the extra-dimension in RS scenario 
\cite{Goldberger:1999uk}. In the GW mechanism, the potential for the radion, 
which determines the size of radius of the extra-dimension can be generated by 
a bulk scalar field with quartic couplings of brane localized potentials. As a 
result, the potential minimum gives a favored compactification scale to solve
 the hierarchy problem. Then a simple extension of the SM in the bulk of the 
warped extra-dimension \cite{Chang:1999nh} and a model with custodial symmetry 
\cite{Agashe:2003zs} have been proposed. 

In this paper, we will focus on a bulk scalar field theory under general 
boundary conditions (BCs) on the warped five-dimension. We analyze profiles of 
vacuum expectation value (VEV) of the scalar under the general BCs. We also 
investigate GW mechanism with the general BCs of the bulk scalar field. We 
point out that the mechanism can work under non-zero Dirichlet BCs that give 
appropriate VEVs. We also consider a scenario that a bulk Higgs field plays a 
role of GW mechanism. Especially, we will show that $SU(2)_R$ triplet Higgs in 
a model with custodial symmetry can be identified with the bulk scalar of GW 
mechanism.

This paper is organized as follows. In section 2, we study behaviors of bulk 
scalar field under possible four BCs on the warped extra-dimension. In section 
3, we investigate the GW mechanism under the BCs in several setups. We will try
 to identify the bulk scalar in the GW mechanism as the Higgs in the bulk SM or
 $SU(2)_R$ triplet Higgs in a model with custodial symmetry. The discussions in
 this section will be proceeded with some reviews of related important models 
and mechanisms. The section 4 is devoted to summary. Relatively technical 
discussions are shown in Appendices.

\section{Bulk scalar in warped extra-dimension}
In this section, we study a bulk scalar field theory on the warped 
extra-dimensional spacetime proposed by Randall and Sundrum, and clarify the 
wave function profiles of classical mode of the scalar under four BCs.

We start with the following action of a bulk field, $\Phi$,
 \begin{eqnarray}
  S=\int d^5x\sqrt{-G}[-G^{MN}(\partial_M\Phi^\dagger)(\partial_N\Phi)
                       -\mathcal{V}(|\Phi|^2)], \label{action}
 \end{eqnarray}
where $x^M=(x^\mu,y)=(x^0,\cdots,x^5)$, $y=x^5$. For simplicity, we assume that
 the potentials solely depend on $|\Phi|^2$ so that potentials can be written 
as $\mathcal{V}(|\Phi|^2)$. The metric is given by
 \begin{eqnarray}
  G_{MN}dx^Mdx^N             = e^{-2\sigma}\eta_{\mu\nu}dx^\mu dx^\nu+dy^2,
  ~~~\mbox{ and }~~~
  G^{MN}\partial_M\partial_N = e^{2\sigma}\eta_{\mu\nu}\partial_\mu
                                 \partial_\nu+\partial_y^2,
 \end{eqnarray}
where 
 \begin{eqnarray}
  \sigma\equiv k|y|,~~~
  \sigma'=k\epsilon(y),~~~
  \sigma''=       2k[\delta(y)-\delta(y-L)],~~~
  \eta_{\mu\nu} \equiv \mbox{diag}\{-1,1,1,1\}.
 \end{eqnarray}
The $\epsilon(y)$ is a kind of sign function defined by $\epsilon(\pm|y|)=\pm1$
 and $\epsilon(0,L)=0$. The $k$ is the brane tension, which is related to the 
bulk energy density (cosmological constant $\Lambda$) and the brane potential 
energy by
 \begin{eqnarray}
  k\equiv\pm\sqrt{\frac{-\Lambda}{6M_5^3}}
  =\frac{V_{\text{UV}}}{6M_5^3}=-\frac{V_{\text{IR}}}{6M_5^3}, 
  \label{brane-tension}
 \end{eqnarray}
where $M_5$ is the Planck mass in five-dimensions. The stable and flat 
configurations of the branes can be realized when the relations 
\eqref{brane-tension} is satisfied. By utilizing the above descriptions, the 
action \eqref{action} can be rewritten by
 \begin{eqnarray}
  S=\int d^4x\int_0^Ldye^{-4\sigma}[-e^{2\sigma}|\partial_\mu\Phi|^2
    -|\partial_y\Phi|^2-\mathcal{V}]. \label{action1}
 \end{eqnarray}
We define the action on a line segment as $0\leq y\leq L$. When we write the 
bulk scalar field as
 \begin{eqnarray}
  \Phi=\frac{\Phi_R+i\Phi_I}{\sqrt{2}},
 \end{eqnarray}
we obtain
 \begin{eqnarray}
  \frac{\partial\mathcal{V}}{\partial\Phi_X}=\Phi_X\mathcal{V}',\hspace{5mm}
  \frac{\partial^2\mathcal{V}}{\partial\Phi_X^2}
  =\mathcal{V}'+\Phi_X^2\mathcal{V}'',\hspace{5mm}
  \frac{\partial^2\mathcal{V}}{\partial\Phi_R\partial\Phi_I}
  =\Phi_R\Phi_I\mathcal{V}'',
 \end{eqnarray}
where $X$ stands for $R$ and $I$, and we have written 
$\mathcal{V}'=d\mathcal{V}/d(|\Phi|^2)$ etc.. The variation of the action is 
given by
 \begin{eqnarray}
  \delta S &=& \int d^4x\int_0^Ldye^{-4\sigma}
                     \bigg[\delta\Phi_X\left(\mathcal{P}\Phi_X
                     -\frac{\partial\mathcal{V}}{\partial\Phi_X}\right) 
               \nonumber \\
           & & \phantom{\int d^4x\int_0^Ldye^{-4\sigma}\bigg[}     
                     +\delta(y)\delta\Phi_X\left(+\partial_y\Phi_X
                     \right)
                     +\delta(y-L)\delta\Phi_X\left(-\partial_y\Phi_X
                     \right)\bigg],
 \end{eqnarray}
where we define as 
$\mathcal{P}=e^{2\sigma}\Box+e^{4\sigma}\partial_ye^{-4\sigma}\partial_y$. The 
VEV of the scalar field is determined by the action principle, $\delta S=0$, 
that is,
 \begin{eqnarray}
  \mathcal{P}\Phi_X-\frac{\partial\mathcal{V}}{\partial\Phi_X}=0, \label{eom}
 \end{eqnarray}
while the BC at $y=0$ and $L$ reads either Dirichlet
 \begin{eqnarray}
  \delta\Phi_X|_{y=\eta}=0 \label{Dirichlet}
 \end{eqnarray}
or Neumann 
 \begin{eqnarray}
  \left.\pm\partial_y\Phi_X\right|_{y=\eta}=0, \label{Neumann}
 \end{eqnarray}
where signs above and below are for $\eta=0$ and $L$, respectively\footnote{If 
one considers a case with brane localized potential, the Neumann type BCs are 
changed. The formulation for the case with brane localized potential is given 
in the Appendix A. For our purpose of this paper, it is enough to discuss in 
the absence of the brane potentials and main results are not modified.}. We can
 have four choices of combination of Dirichlet and Neumann BCs at $y=0$ and 
$L$, namely $(D,D)$, $(D,N)$, $(N,D)$, and $(N,N)$. Different choice of BC 
corresponds to different choice of the theory. Once the theory is fixed, one of
 the four conditions is determined.

We study behaviors of the bulk scalar field on the warped five dimension by 
utilizing the background field method, separating the field into classical and 
quantum fluctuation parts:
 \begin{eqnarray}
  \Phi(x,y)=\Phi^c(x,y)+\phi^q(x,y).
 \end{eqnarray}
The configuration of the classical field obeys the EOM \eqref{eom},
 \begin{eqnarray}
  \mathcal{P}\Phi_X^c-\frac{\partial\mathcal{V}}{\partial\Phi_X}^c=0, 
 \end{eqnarray}
with either the Dirichlet BC
 \begin{eqnarray}
  \delta\Phi_X^c|_{y=\eta}=0,
 \end{eqnarray}
or the Neumann BC
 \begin{eqnarray}
  \left.\pm\partial_y\Phi_X^c\right|_{y=\eta}=0,
 \end{eqnarray}
at each brane\footnote{The additional effects induced by the brane terms also 
change the VEV and quantum field wave-function profiles, which lead to 
interesting phenomenological consequences \cite{Haba:2009uu,HOT3}.}. Here and 
hereafter, we use the following shorthand notation,
 \begin{eqnarray}
  \frac{\partial V}{\partial\Phi}^c(x,y)
  \equiv\left.\frac{\partial V}{\partial\Phi}\right|_{\Phi=\Phi^c(x,y)},
  \hspace{5mm}\frac{\partial^2 V}{\partial\Phi^2}^c(x,y)
  \equiv\left.\frac{\partial^2V}{\partial\Phi^2}\right|_{\Phi=\Phi^c(x,y)},
 \end{eqnarray}
etc..

For simplicity, we take the bulk potential as 
 \begin{eqnarray}
  \mathcal{V}=m^2|\Phi|^2=\frac{m^2}{2}(\Phi_R^2+\Phi_I^2). \label{bulk-pot}
 \end{eqnarray}
In this case, the EOM \eqref{eom} can be written down as 
 \begin{eqnarray}
  (\partial_y^2-4k\partial_y-m^2)\Phi_X^c=0.
 \end{eqnarray}
The solution of this equation is given by
 \begin{eqnarray}
  \Phi_X^c(z)=Az^{\nu+2}+Bz^{-(\nu-2)}, \label{general}
 \end{eqnarray}
where $\nu\equiv\sqrt{4+m^2/k^2}$ and $z\equiv e^\sigma$. 

Next, let us study the profile of quantum fluctuation of the scalar field. We 
separate the field into the classical field and quantum fluctuation as
 \begin{eqnarray}
  \Phi(x,y) &=& \frac{1}{\sqrt{2}}\left[v(y)+\phi(x,y)+i\chi(x,y)\right],
                \label{expansion} \\
            &=& \frac{1}{\sqrt{2}}
                \left[v(y)+\sum_{n=0}^{\infty}f_n^\phi(y)\phi_n(x)
                      +i\sum_{n=0}^{\infty}f_n^\chi(y)\chi_n^q(x)\right],
 \end{eqnarray}
where we took $\Phi_R=v(y)$ and $\Phi_I=0$, and the Kaluza-Klein (KK) 
expansions are taken for $\phi(x,y)$ and $\chi(x,y)$. We put separation 
\eqref{expansion} into \eqref{action1} and expand up to the quadratic terms of 
the field $\phi$ and $\chi$ as\footnote{The derivation of the action with brane
 localized potentials is given in Appendix A.} 
 \begin{eqnarray}
  S(\phi,\chi) &=& \int d^4x\int_0^Ldye^{-4\sigma}
        \Bigg[\frac{1}{2}\phi
              \Bigg(e^{2\sigma}\Box
                    +e^{4\sigma}\partial_ye^{-4\sigma}\partial_y
                    -\frac{\partial^2\mathcal{V}}{\partial\Phi_R^2}^c\Bigg)\phi
        \nonumber \\
    & & +\frac{1}{2}\chi
         \Bigg(e^{2\sigma}\Box
               +e^{4\sigma}\partial_ye^{-4\sigma}\partial_y
               -\frac{\partial^2\mathcal{V}}{\partial\Phi_I^2}^c\Bigg)\chi
        \nonumber \\
    & & -\frac{\delta(y)}{2}
         (-\phi\partial_y\phi-\chi\partial_y\chi) 
         -\frac{\delta(y-L)}{2}
         (+\phi\partial_y\phi+\chi\partial_y\chi)\Bigg].
 \end{eqnarray}
This corresponds to the bulk action for $\phi$. By utilizing KK expansion, the 
KK equation is given by\footnote{We focus only on the $\phi$ field throughout 
this paper unless it is needed to distinguish between the $\phi$ and $\chi$ 
fields.}
 \begin{eqnarray}
  e^{-4\sigma}\Bigg(\partial_y^2-4k\partial_y
        -\frac{\partial^2\mathcal{V}}{\partial\Phi_R^2}^c\Bigg)f_n^\phi(y)
  =-\mu_{\phi_n}^2f_n^\phi(y).
 \end{eqnarray}
The general Dirichlet and Neumann BCs are
 \begin{eqnarray}
  f_n^\phi(y)|_{y=\eta}=0,
 \end{eqnarray}
and 
 \begin{eqnarray}
  \pm\partial_yf_n(y)|_{y=\eta}=0, \label{Neumann-quantum}
 \end{eqnarray}
respectively. 
In this setup, we investigate the profile of bulk scalar under the above four 
BCs.

\subsection{{\boldmath $(D,D)$} case}
\label{sec-DD}

First, we study a case in which both BCs on the $y=0$ and $y=L$ branes, which 
correspond to $z=1$ and $z=e^{kL}\equiv z_L$ ones respectively, are the 
Dirichlet type BCs. The most general form of the Dirichlet BC is 
$\delta\Phi|_{z=\xi}=0$ and 
 \begin{eqnarray}
  v(1) = v_1,~~~ 
  v(z_L) = v_2, \label{DD-L}
 \end{eqnarray}
where $\xi$ is taken as $1$ and $z_L$. These BCs can be rewritten as
 \begin{eqnarray}
  A+B = v_1,~~~ 
  Az_L^{\nu+2}+Bz_L^{-(\nu-2)} = v_2, \label{DD-L-1}
 \end{eqnarray}
by utilizing the general solution \eqref{general} of the EOM. The 
\eqref{DD-L-1} lead to
 \begin{eqnarray}
  A = -\frac{v_1z_L^{-(\nu-2)}-v_2}{z_L^{\nu+2}-z_L^{-(\nu-2)}},~~~ 
  B = \frac{v_1z_L^{\nu+2}-v_2}{z_L^{\nu+2}-z_L^{-(\nu-2)}}. \label{DD-B}
 \end{eqnarray}
Therefore, under these BCs, we obtain the VEV profile as
 \begin{eqnarray}
  v(z)=-\frac{v_1z_L^{-(\nu-2)}-v_2}{z_L^{\nu+2}-z_L^{-(\nu-2)}}z^{\nu+2}
            +\frac{v_1z_L^{\nu+2}-v_2}{z_L^{\nu+2}-z_L^{-(\nu-2)}}z^{-(\nu-2)}.
 \end{eqnarray}
A typical profile is shown in Tab. \ref{tab1}. It is seen that the VEV profile 
localizes toward to the IR brane. In order to solve the hierarchy problem in 
the RS background, the magnitude of $z_L$ becomes $\mathcal{O}(10^{15})$. 

\subsection{{\boldmath$(D,N)$} case}

Next, let us consider the $(D,N)$ case. These BCs can be described as
 \begin{eqnarray}
  v(1)=v_1,~~~\partial_zv(z)|_{z=z_L}=0.
 \end{eqnarray}
Then, they are written down by
 \begin{eqnarray}
  A+B=v_1,~~~A(\nu+2)z_L^{\nu+2}-B(\nu-2)z_L^{-(\nu-2)}=0.
 \end{eqnarray}
These lead to 
 \begin{eqnarray}
  v(z)=\frac{v_1(\nu-2)z_L^{-(\nu-2)}}
            {(\nu+2)z_L^{\nu+2}+(\nu-2)z_L^{-(\nu-2)}}z^{\nu+2}
       +\frac{v_1(\nu+2)z_L^{\nu+2}}
            {(\nu+2)z_L^{\nu+2}+(\nu-2)z_L^{-(\nu-2)}}z^{-(\nu-2)}.
 \end{eqnarray}
The profile is illustrated in  Tab. \ref{tab1}.
 
\subsection{{\boldmath$(N,D)$ case}}

In $(N,D)$ BC case, the BCs are
 \begin{eqnarray}
  \partial_z v(z)|_{z=1}=0,~~~v(z_L) = v_2.
 \end{eqnarray}
And they can be written down as
 \begin{eqnarray}
  A(\nu+2)-B(\nu-2)=0,~~~ Az_L^{\nu+2}+Bz_L^{-(\nu-2)}=v_2. \label{BC-ND-2}
 \end{eqnarray}
These lead to
 \begin{eqnarray}
  v(z)=\frac{(\nu-2)v_2}{(\nu-2)z_L^{\nu+2}+(\nu+2)z_L^{-(\nu-2)}}z^{\nu+2}
       +\frac{(\nu+2)v_2}{(\nu-2)z_L^{\nu+2}+(\nu+2)z_L^{-(\nu-2)}}
        z^{-(\nu-2)}.
 \end{eqnarray}
The profile is shown in Tab. \ref{tab1}.

\subsection{{\boldmath$(N,N)$ case}}
\label{sec-NN}

Finally, we discuss the $(N,N)$ BC case. The BCs are  
 \begin{eqnarray}
  \partial_z v(z)|_{z=1}=0,~~~\partial_zv(z)|_{z=z_L}=0,
 \end{eqnarray}
and they are written down as
 \begin{eqnarray}
  A[(\nu+2)-B(\nu-2)]=0,~~~A(\nu+2)z_L^{\nu+2}-B(\nu-2)z_L^{-(\nu-2)}=0.
 \end{eqnarray} 
We find that there is no solution to satisfy the above BCs except for a trivial
 one, $(A,B)=(0,0)$, which might not have physical interests in any 
phenomenological models. That does not depends on whether the brane localized 
potentials exist or not, that is, there is no solution, which is consistent 
with the BC, if the brane localized potentials exist. Therefore, it is not 
trivial that there is a viable VEV of a bulk scalar field satisfying the 
$(N,N)$ type BCs in any phenomenological models, which would generally have 
brane localized interactions from radiative collections. 

In this section, we formulated a scalar field theory under four BCs on a warped
 five-dimensional background. Then a VEV profile of the bulk scalar field was 
given by both analytic and numerical computations. It is straightforward to 
extend the above discussions to a higher extra-dimensional background. In the 
next section, we study some applications of role of bulk scalar field in warped
 five-dimensional models.

\section{Warped five-dimensional models with bulk scalar}

We investigate some applications of role of bulk scalar field, whose VEV 
profile is presented in the previous section, in warped five-dimensional 
models. Especially, following applications are (re)considered, (i) realizations
 of the GW mechanism under general BCs discussed in the previous section, (ii) 
the bulk SM Higgs as the GW scalar, and (iii) a triplet Higgs under additional 
$SU(2)_R$ symmetry as the GW scalar.

\subsection{GW mechanism and general BCs}

In this section, we discuss the GW mechanism \cite{Goldberger:1999uk}, where a 
bulk scalar plays an important role to stabilize the radion. We give a short 
review of this mechanism at first.

Goldberger and Wise proposed a mechanism for stabilizing the size of 
extra-dimension in the warped space. The GW mechanism starts with the bulk and 
brane actions for a bulk scalar field given in \eqref{action1-app}, 
\eqref{bulk-pot}, and \eqref{brane-pot}. The VEV of the bulk scalar field can 
be classically obtained by solving the EOM, and its general solution is given 
in \eqref{general}. The unknown coefficients $A$ and $B$ are determined by 
imposing BCs as we performed in the section \ref{sec-DD}-\ref{sec-NN}. In the 
GW mechanism, the $(N,N)$ type BCs have been taken. And the mechanism includes 
the brane localized potentials. Therefore, the BCs can be written down as 
 \begin{eqnarray}
  && k[A(\nu+2)-B(\nu-2)]-\lambda_0(A+B)[(A+B)^2-v_0^2]=0, \\
  && k[A(\nu+2)z_L^{\nu+2}-B(\nu-2)z_L^{-(\nu-2)}] \nonumber \\
  && +\lambda_L(Az_L^{\nu+2}+Bz_L^{-(\nu-2)})
      [(Az_L^{\nu+2}+Bz_L^{-(\nu-2)})^2-v_L^2]=0.
 \end{eqnarray} 
However, the mechanism is considered in a case where the boundary quartic 
couplings $\lambda_0$ and $\lambda_L$ are large. Here, one should note that the
 $(N,N)$ BCs in the limit of large boundary couplings with the potential 
\eqref{bulk-pot} and \eqref{brane-pot} are equivalent to the $(D,D)$ ones given
 in \eqref{DD-L} if we take $v_1=v_0$ and $v_2=v_L$. Since $A$ and $B$ in the 
GW mechanism with the large coupling limits and in expansion of $z_L^{-1}$ are 
actually given by
 \begin{eqnarray}
  A_{\text{GW}} &\simeq& -v_0z_L^{-2\nu}+v_Lz_L^{-(\nu+2)}, \\
  B_{\text{GW}} &\simeq& v_0(1+z_L^{-2\nu})-v_Lz_L^{-(\nu+2)},
 \end{eqnarray}
it is easily seen that $A$ and $B$ given in \eqref{DD-B} are 
$A\simeq A_{\text{GW}}$ and $B\simeq B_{\text{GW}}$ under the limits and 
replacements of $v_1=v_0$ and $v_2=v_L$. The above correspondence from the 
$(N,N)$ BCs with large boundary couplings to $(D,D)$ can be also simply 
understood in terms of the vacuum structure for an effective four-dimensional 
potential. The effective potential in four dimensions are written down by 
putting \eqref{general} back into \eqref{action1-app} and integrating over the 
extra-dimension as
 \begin{eqnarray}
  V_{\text{eff}} &=& k(\nu+2)A^2(z_L^{2\nu}-1)+k(\nu-2)B^2(1-z_L^{-2\nu}) 
                     \nonumber \\
                 & & +\lambda_Lz_L^{-4}[\Phi^c(z_L)^2-v_L^2]^2
                     +\lambda_0[\Phi^c(1)^2-v_0]^2.
 \end{eqnarray}
We find that $\Phi^c(1)=v_0$ and $\Phi^c(z_L)=v_L$ are energetically favored at
 the large boundary coupling limit, and thus, these solutions just correspond 
to the $(D,D)$ BCs in \eqref{DD-L} with $v_1=v_0$ and $v_2=v_L$.

In \cite{Goldberger:1999uk}, it was assumed for simplicity that
 \begin{eqnarray}
  \frac{m}{k}\ll1,
 \end{eqnarray}
so that 
 \begin{eqnarray}
  \nu=2+\epsilon~~~\mbox{ with }~~~\epsilon\simeq\frac{m^2}{4k^2}.
 \end{eqnarray}
Under this assumption, the effective potential becomes
 \begin{eqnarray}
  V_{\text{eff}}=k\epsilon v_0^2
                 +4kz_L^{-4}(v_L-v_0z_L^{-\epsilon})^2
                  \left(1+\frac{\epsilon}{4}\right)
                 -k\epsilon v_0z_L^{-(4+\epsilon)}(2v_L-v_0z_L^{-\epsilon}).
 \end{eqnarray}
{}For the purpose of moduli stabilization in GW mechanism, we rewrite the 
potential as
 \begin{eqnarray}
  V_{\text{eff}}=k\epsilon v_0^2+4ke^{-4kL}(v_L-v_0e^{-\epsilon kL})^2
                                 \left(1+\frac{\epsilon}{4}\right)
                 -k\epsilon v_0e^{-(4+\epsilon)kL}(2v_L-v_0e^{-\epsilon kL}),
  \label{radion-pot}
 \end{eqnarray}
where $z_L=e^{kL}$ is utilized. This potential can be approximately minimized 
at 
 \begin{eqnarray}
  kr_c\equiv\frac{kL}{\pi}
      \simeq\left(\frac{4}{\pi}\right)\frac{k^2}{m_2}\ln
            \left[\frac{v_0}{v_L}\right]. 
 \end{eqnarray}
It is seen that the mechanism requires only 
 \begin{eqnarray}
  \frac{m^2}{k^2}\simeq\mathcal{O}(10),
 \end{eqnarray}
in order to realize $kr_c\sim10$, which is needed for solving the hierarchy 
problem in RS background. Therefore, it can be concluded that fine-tuning among
 parameters is not required to stabilize the configuration of radion which can 
play a crucial role in this approach for the gauge hierarchy problem. Finally,
 a numerical example is given by
 \begin{eqnarray}
  \frac{v_0}{v_L}=1.5,~~~\frac{m}{k}=0.2,~~~kr_c=12. \label{example}
 \end{eqnarray}

The effects of the radion on the oblique parameters, $S$, $T$, and $U$ 
\cite{stu1,stu2,stu3}, have been evaluated by using an effective theory 
approach in \cite{Csaki:2000zn}. As shown in 
\cite{Csaki:2000zn}, in the absence of a curvature-scalar Higgs mixing operator
 such as $\xi\mathcal{R}H^\dagger H$, the magnitude of the contribution to the 
oblique parameters from the radion can be small.
On the other hand, in the presence of the mixing operator,
the corrections become large due to the modified radion-Higgs couplings.
As the results, the magnitude of fine-tuning among model parameters should be 
increased to achieve the Higgs mass larger than a few hundred GeV. 
Therefore, there are two options to protect the oblique 
parameters within an experimentally allowed region. One is to assume the 
absence of the curvature-scalar Higgs mixing operator. The other is to tune the
 parameters $\xi$ and $v_{\text{EW}}/(M_{pl}e^{-kr_c})$ so that the oblique 
parameter are within an allowed region (see the Fig.~5 in \cite{Csaki:2000zn} 
for an allowed region of $\xi$ and $v_{\text{EW}}/(M_{pl}e^{-kr_c})$ given by 
a numerical calculation). The above options to control effects from radion can 
be generically taken for actual models discussed in the following subsections.

\begin{figure}
\begin{center}
\begin{tabular}{cccc}
 & $\tilde{\lambda}_L=1$ & $\tilde{\lambda}_L=5$ & $\tilde{\lambda}_L=10^2$ \\
{}\raisebox{1.3cm}{[$(D,N)$ case]} & \includegraphics[scale = 0.42]{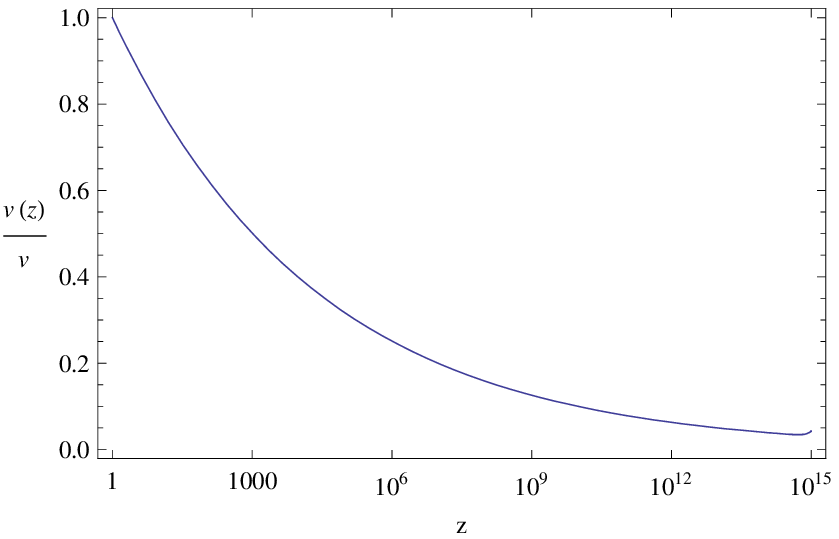} & \includegraphics[scale = 0.42]{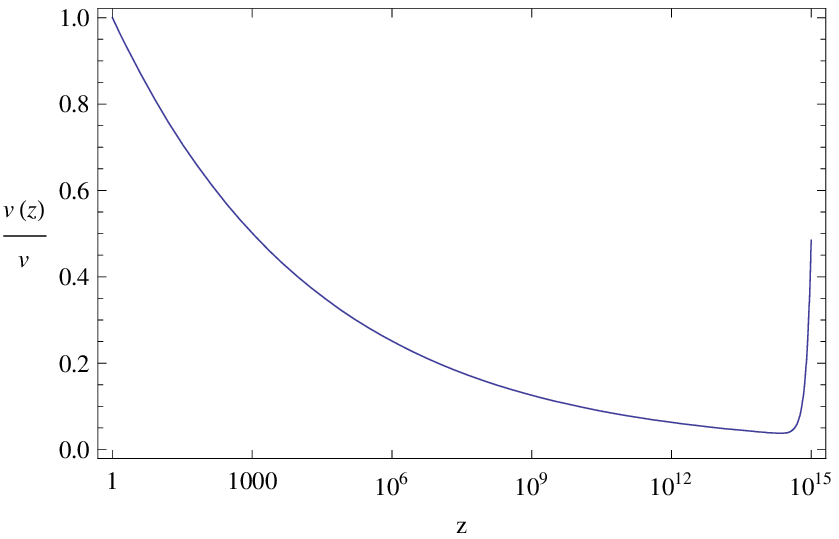} & \includegraphics[scale = 0.42]{fig-app-3-100.eps} \\
 & $\tilde{\lambda}_0=0.01$ & $\tilde{\lambda}_0=0.2$ & $\tilde{\lambda}_0=10^2$ \\
{}\raisebox{1.3cm}{[$(N,D)$ case]} & \includegraphics[scale = 0.42]{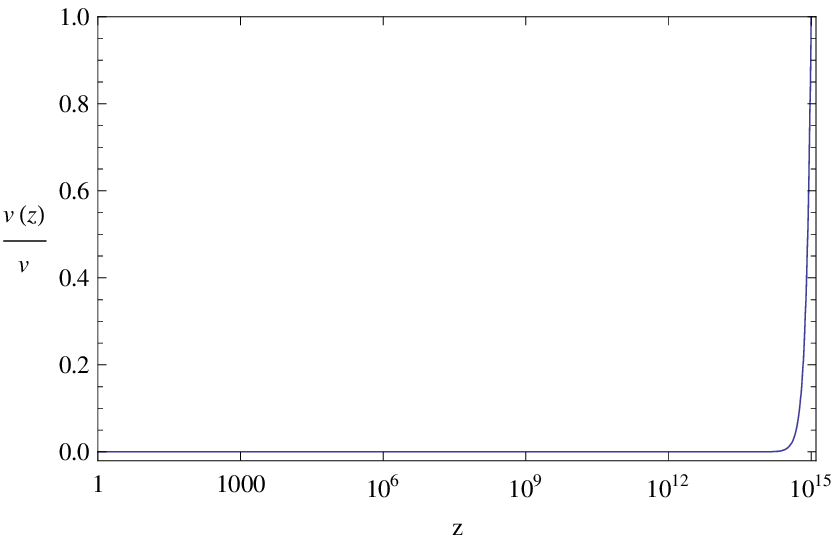} & \includegraphics[scale = 0.42]{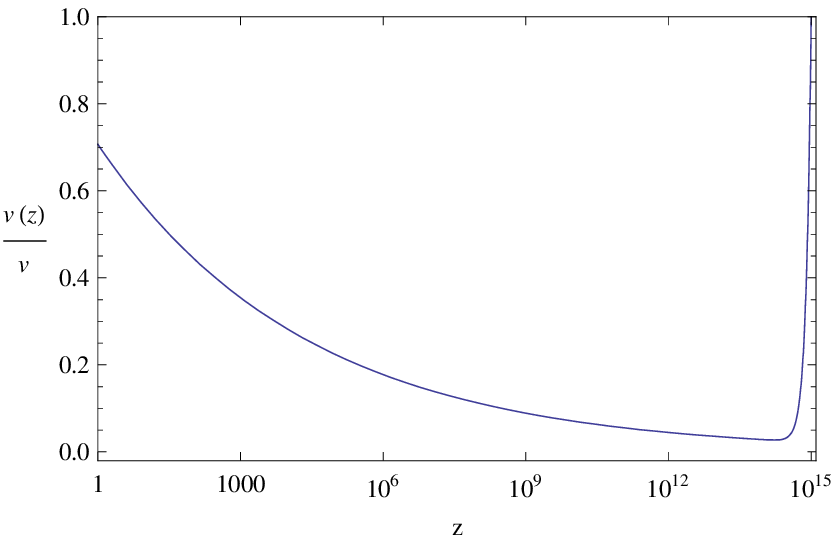} & \includegraphics[scale = 0.42]{fig7-2-100.eps}
\end{tabular}
\end{center}
\caption{Dependence of VEV profile on the boundary coupling. We take the same 
values for other parameters as ones in Tab. \ref{tab1}.}
\label{fig0}
\end{figure}

At the end of this subsection, it is worth commenting on realizations of the 
mechanism itself under various BCs discussed in the previous section. Since the
 Neumann BC including the effect from brane potential with huge boundary 
quartic coupling, which is imposed in the GW mechanism, is equivalent to the 
Dirichlet BC, the GW mechanism can work in the $(D,D)$ BCs case with 
appropriate VEVs. Moreover, the mechanism can be also realized in $(D,N)$ and 
$(N,D)$ BCs including the effect from brane potential with large boundary 
quartic coupling while it cannot work in the absence of brane potentials. These
 are summarized in Tab \ref{tab1}. We also analyze  the VEV profile when brane 
localized potentials are introduced. The Fig. \ref{fig0} shows the VEV profile 
with the boundary quartic coupling. The drastical change of VEV profile in the 
$(D,N)$ case can be seen when imposing larger boundary coupling, 
$\tilde{\lambda}_L\gtrsim\mathcal{O}(1)$. This significant change compared to 
the $(N,D)$ case, $\tilde{\lambda}_0\gtrsim\mathcal{O}(0.1)$ is explained as 
follows: The VEV profiles of both $(D,D)$ and $(N,D)$ cases in the absence of 
brane potential sharply localized on the IR brane while the profile of $(D,N)$ 
case gently localized on the UV brane. potential), The required potential 
energy at the UV brane which changes the VEV profile of $(N,D)$ case is smaller
 than one at IR boundary. It should drastically change the profile of $(D,N)$ 
case. In other word, the form of VEV profile in the $(N,D)$ case is more 
sensitive to the effect of brane potential than $(D,N)$ or $(D, D)$ cases. As 
for $(N,N)$ case with large boundary potential, it is just the case of original
 GW mechanism, and it has been shown to work well \cite{Goldberger:1999uk}.
\begin{table}
\begin{center}
\begin{tabular}{c|c|c|c}
\hline
\hline
\multicolumn{4}{c}{VEV profiles under various BCs in the absence of brane localized potentials}\\
\hline
\hline
$(D,D)$ & $(D,N)$ & $(N,D)$ & $(N,N)$ \\
\hline
\includegraphics[scale = 0.42]{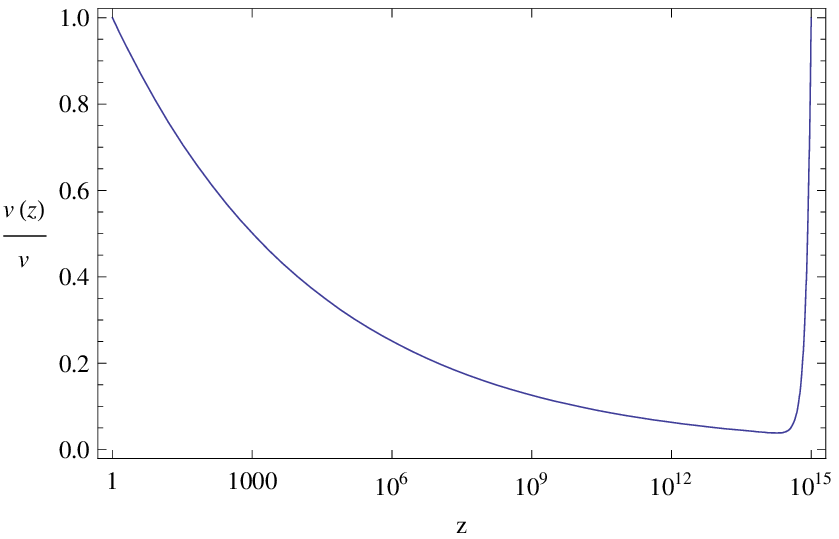} & \includegraphics[scale = 0.42,bb=0 0 240 160]{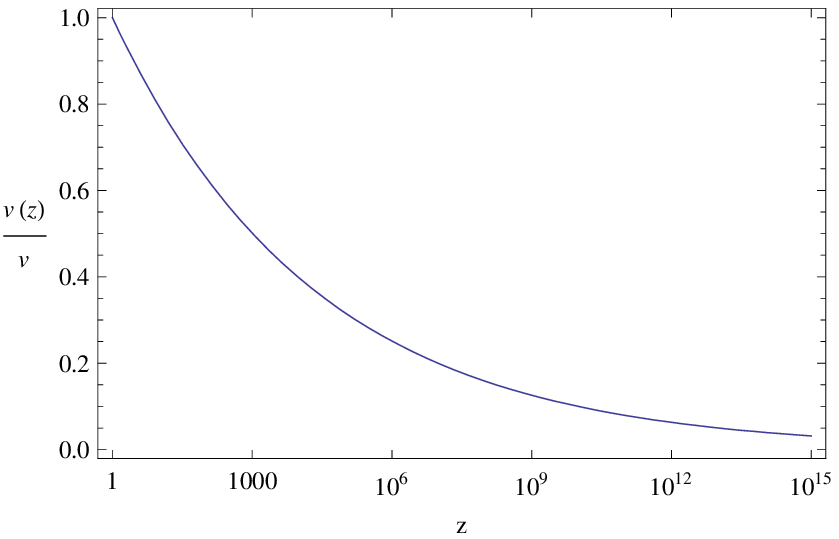} & \includegraphics[scale = 0.42]{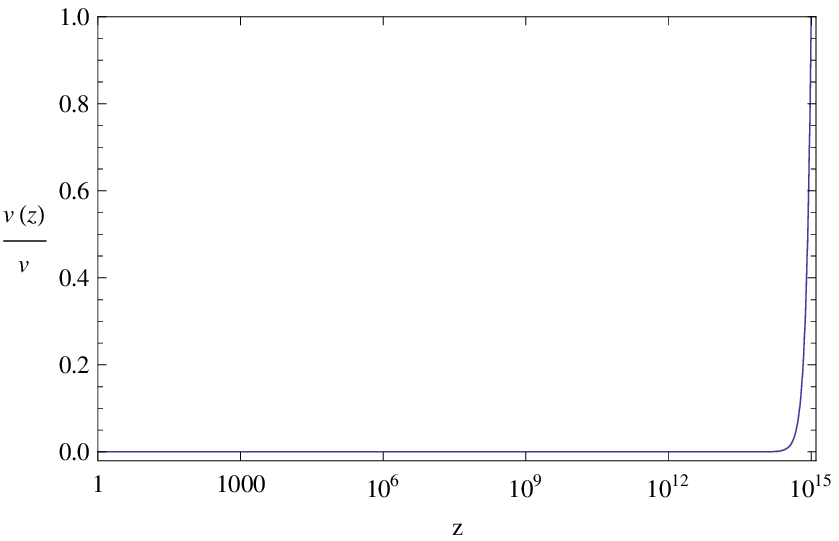} & \includegraphics[scale = 0.42]{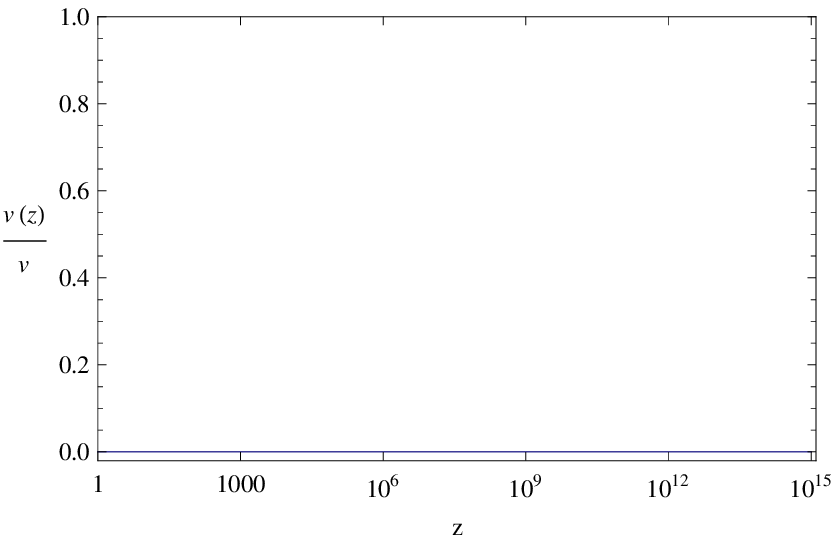} \\
\hline
\hline
\multicolumn{4}{c}{VEV profiles under various BCs in the presence of brane localized potentials}\\
\hline
\hline
$(D,D)$ & $(D,N)$ & $(N,D)$ & $(N,N)$ \\
\hline
\includegraphics[scale = 0.42]{fig1-2.eps} & \includegraphics[scale = 0.42]{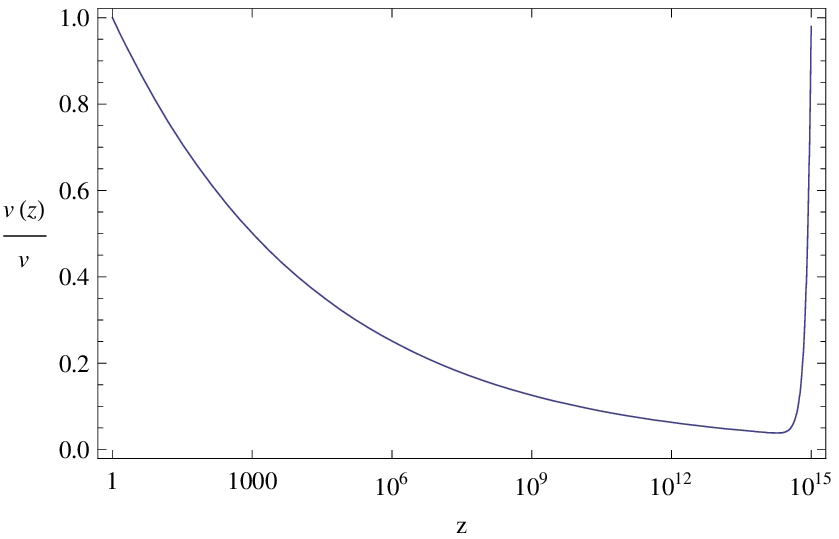} & \includegraphics[scale = 0.42]{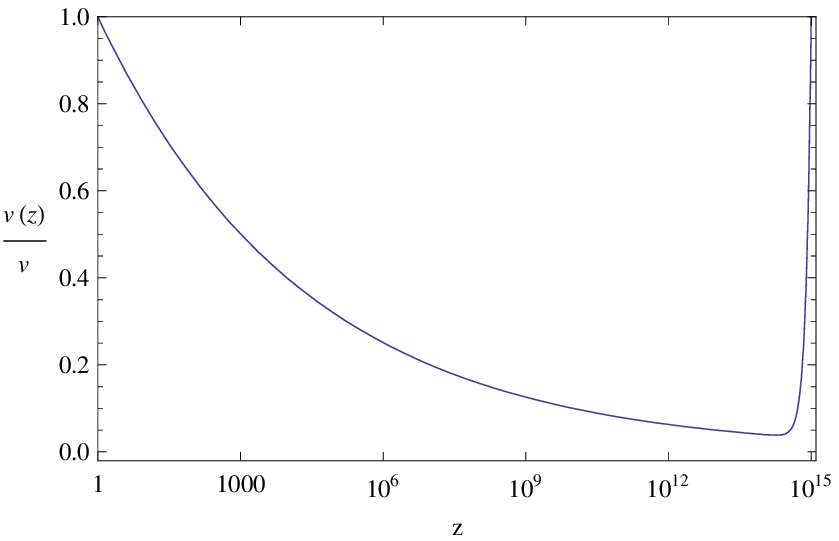} & \includegraphics[scale = 0.42]{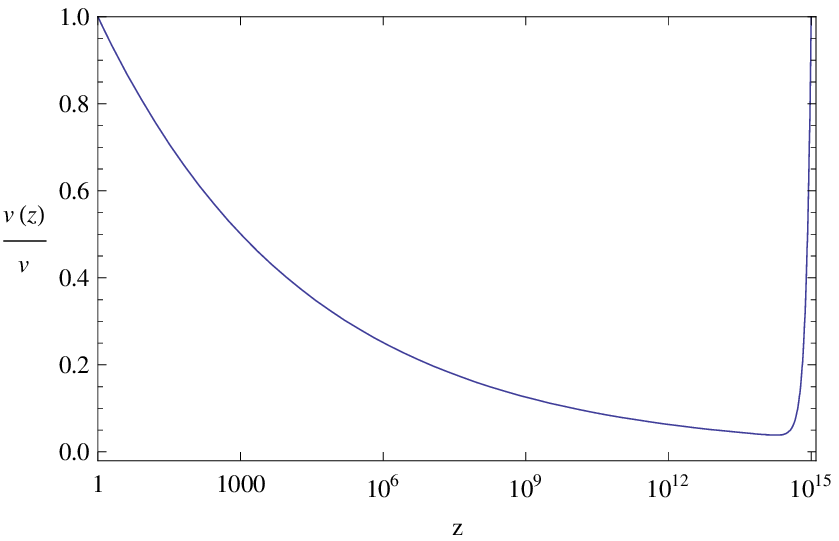} \\
\hline
\hline
\multicolumn{4}{c}{Realization of the GW mechanism}\\
\hline
\hline
$(D,D)$ & $(D,N)$ & $(N,D)$ & $(N,N)$ \\
\hline
$\bigcirc$ & $\bigcirc^\ast$ & $\bigcirc^\ast$ & $\bigcirc^\ast$ \\
\hline
\hline
\end{tabular}
\end{center}
\caption{VEV profiles and realization of GW mechanism: We redefine all 
dimension-full parameters as 
$v\equiv v_{0,1,2,L}\equiv\tilde{v}_{0,1,2,L}M_{\text{pl}}^{3/2}$, 
$\lambda_{0,L}\equiv\tilde{\lambda}_{0,L}M_{\text{pl}}^{-2}$ and  
$k\equiv\tilde{k}M_{\text{pl}}$, and take dimensionless parameters as 
$\tilde{v}_{0,1,2,L}=\tilde{k}=1$ and $\nu=2.1$, and the Planck scale as 
$M_{\text{pl}}=2.4\times10^{18}$ GeV. Notice that the bulk scalar is 
canonically normalized at the UV brane and hence the values of VEV at the IR 
brane is one for unnormalized scalar field, which should be canonically 
normalized for the four-dimensional effective theory later. In the upper 
figures, the rapid changes of profiles near the boundary $z=z_L$ in the $(D,D)$
 and $(N,D)$ cases are due to the Dirichlet BCs at $z=z_L$. In the $(D,N)$ case
 of upper figure, the VEV at IR brane becomes smaller than Planck scale but is 
still finite, $v(z_L)/v\sim0.03$. In the lower figures, we take the brane 
localized potential as given in \eqref{brane-pot} with large boundary 
couplings, $\tilde{\lambda}_{0,L}=10^2$, for the $(D,N)$ and $(N,D)$ cases, and
 utilize the approximated solution of EOM at huge boundary coupling limit, 
$\tilde{\lambda}_{0,L}\gg1$, for the $(N,N)$ case which is presented in the GW 
mechanism \cite{Goldberger:1999uk}. It can be seen that the Neumann type BC 
with brane localized potential including a huge quartic coupling becomes 
equivalent to the non-vanishing Dirichlet BC. $\bigcirc^\ast$ means that the GW
 mechanism can work in the case that Neumann BC includes effects of brane 
localized potentials with huge boundary quartic coupling. The figure for 
$(D,D)$ case without the brane localized potentials is the same as one for 
$(D,D)$ case with brane potentials.} 
\label{tab1}
\end{table}

\subsection{GW mechanism with bulk SM Higgs}

In this subsection, we study a possibility of bulk SM Higgs with brane 
fermions, which is one of simple extensions of the SM. The constraints on the 
KK scale, $m_{KK}=\pi/L$ in flat five-dimensional spacetime from the EW 
precision measurements have been discussed in refs. 
\cite{Nath:1999fs,Masip:1999mk,Rizzo:1999br,Strumia:1999jm,Carone:1999nz}: 
$m_{KK}>1.7$ TeV ($90\%$ CL) from the experimentally observed value of the 
Fermi constant \cite{Nath:1999fs}, $m_{KK}\gtrsim2.5$ TeV from the leptonic $Z$
 width \cite{Masip:1999mk}, $m_{KK}\gtrsim3.8$ TeV ($95\%$ CL) from a global 
fit of measurements of the Fermi constant, $\Gamma(Z\rightarrow f\bar{f})$, 
atomic parity violation, Weinberg angle, and $W$ boson mass etc. when $m_H<260$
 GeV \cite{Rizzo:1999br}, $m_{KK}>3.5$ TeV from a global fit of Fermi constant,
 $Z$, $W$, top  masses, $Z$ widths, asymmetries in $Z$ 
decays\footnote{$m_{KK}>4.3$ TeV if the Higgs is confined to brane 
\cite{Strumia:1999jm}.} \cite{Strumia:1999jm}, $m_{KK}>85$ TeV from 
$K$-$\bar{K}$ and $D$-$\bar{D}$ mixing in bulk generation scenario where the 
first two generations live in the bulk together with the gauge multiplets and 
one of two Higgs fields \cite{Carone:1999nz}, $m_{KK}>1.52$ TeV ($95\%$ CL) 
from the measurement of the Fermi constant in $SU(2)$-brane scenario 
\cite{Carone:1999cb,Delgado:1999ba} from the $Z$ leptonic width 
\cite{Carone:1999nz}. On the other hand, there is a phenomenologically 
interesting predictions in addition to the presence of KK particles in a 
higher-dimensional model, which is the {\it Yukawa deviation} 
\cite{Haba:2009pb,Haba:2009uu}. The Yukawa deviation is a phenomenon that the 
Yukawa coupling is smaller than the naive SM expectation, i.e. the SM fermion 
mass divided by the Higgs VEV. Such deviation can generally occur in 
multi-Higgs models, e.g. minimal supersymmetric standard model (MSSM). However,
 it has been pointed out that the Yukawa deviation can be induced from the 
presence of brane localized Higgs potential, which leads to deformed 
wave-function profile in the bulk for zero-mode physical Higgs, in 
extra-dimensional setup even if there exists only one Higgs doublet 
\cite{Haba:2009uu}. Furthermore, the Dirichlet Higgs model where 
extra-dimensional BCs are Dirichlet type predicts the maximal Yukawa deviation
 with brane localized SM fermions \cite{Haba:2009pb}. How about a reliability 
of the Yukawa deviation in warped extra-dimension? We have shown the VEV 
profile of a bulk scalar field in all cases of BCs, $(D,D)$, $(D,N)$, $(N,D)$, 
and $(N,N)$. As shown in the previous section, the VEV profile localizes toward
 to the UV brane due to the Dirichlet BC in the $(D,N)$ case. Such kind of 
model will be generically problematic when the SM gauge fields are in the bulk 
and the zero mode of bulk scalar field is identified with the SM Higgs boson. 
Since the gauge boson masses can be obtained from\footnote{A detailed deviation
 is given in the Appendix \ref{B}.}
 \begin{eqnarray}
  S &=& -\int d^4x\int_0^Ldye^{-2\sigma}
        \left[\frac{e^2}{4s_W}v(y)^2W_\mu^+(x,y)W^{-\mu}(x,y)\right. 
        \nonumber\\
    & & \phantom{-\int d^4x\int_0^Ldye^{-2\sigma}\bigg[}
        \left.
        +\frac{e^2}{2(\sin2\theta_W)^2}v(y)^2Z_\mu (x,y)Z^\mu(x,y)\right],
  \label{gauge-mass}
 \end{eqnarray}
bulk masses depend on the VEV profile. If the profile localizes toward to the 
UV brane, the realistic values of gauge boson masses cannot be reproduced at 
the IR brane. {}For the $(D,D)$ and $(N,D)$ BC cases, the VEV profiles localize
 toward to the IR brane. Therefore, SM gauge boson masses would be realized to 
be the same as the SM for $W$ and $Z$ bosons if effects from the bulk mass 
could be enough small. However, there are not solutions of the KK equation for 
the physical Higgs (quantum) field because of the presence of warp factor and 
Dirichlet BC at IR brane unlike the case of flat extra-dimension 
\cite{Haba:2009pb,Haba:2010xz}. Finally, $(N,N)$ case cannot lead to 
non-trivial solutions of EOM (VEV). We can now conclude that the Yukawa 
deviation cannot occur in a realistic warped extra-dimensional model with brane
 fermions, bulk Higgs and gauge bosons even when there exists the 
brane-localized Higgs potential unlike a flat extra-dimension model 
\cite{Haba:2009pb,Haba:2009uu,Haba:2010xz}.

Next, we discuss models with the bulk SM Higgs and fermions. One of important 
models with the bulk SM field in a flat five-dimensional spacetime is the UED 
\cite{Appelquist:2000nn}. In the work, it has been pointed out that the bound 
on the size of extra-dimension is weakened compared with other models including
 brane SM fermions and generations etc. due to the presence of KK parity that 
is consistent with an assumption that all SM fields live in the bulk. 
Furthermore, such kind of parity can also make the lightest KK particle a 
candidate for DM. The lower bound on the KK scale from the EW precision 
measurements is $m_{KK}\gtrsim250$ GeV in the originally proposed UED model 
when the Higgs mass is relatively heavy as $m_H\simeq950$ GeV ($90\%$ CL) 
\cite{Appelquist:2000nn,Gogoladze:2006br}. More general setup of the UED models
 have been discussed in \cite{Haba:2009uu,HOT3,Flacke:2008ne}. The ref. 
\cite{Haba:2009uu} has also studied bulk fermion scenario in a case that the 
brane-localized potentials are introduced. Such kind of setup can also lead to 
the Yukawa deviation as discussed above. However, in order to realize a 
detectable size of the deviation at the LHC experiment, the relatively strong 
coupling is needed in the model with bulk SM fermions. The work \cite{HOT3} 
gave the complete computations of the KK expansion of the Higgs and gauge 
bosons in the UED model with brane localized potential by treating the 
potential as a small perturbation, and checked that the $\rho$ parameter is not
 altered by effects from the potential. An alternatively generalized model from
 the UED \cite{Flacke:2008ne} have analyzed effects from the existence of the 
brane localized kinetic and mass terms upon the extra-dimensional 
wave-functions profiles. A different approach to break EW symmetry from the UED
 is the Dirichlet Higgs model \cite{Haba:2009pb,Haba:2010xz,Oda:2010te}. The 
model has a different structure of the Higgs sector from that of the UED, that 
is, the gauge symmetry is broken by non-zero Dirichlet BCs on the bulk Higgs 
field, and there are not any quartic interactions. As the results of this 
setup, the zero mode of the Higgs disappears and its lowest (first) KK mode 
couples to the zero modes of other SM fields with a suppression factor 
$2\sqrt{2}/\pi\simeq0.9$. The detailed phenomenological aspects of this model 
for the LHC experiments have been discussed in \cite{Haba:2010xz}. The most 
important prediction of this model is that the physical Higgs mass is equal to 
the KK scale, $m_H=m_{KK}$, and the current EW precision measurements limit the
 mass to $430\mbox{ GeV}\lesssim m_H\lesssim500$ GeV. The ref. 
\cite{Oda:2010te} has pointed out that the $\mathcal{O}(\sqrt{s})$ growth of 
the longitudinal $W^+W^-$ elastic scattering amplitude is exactly cancelled and
 hence can be uniterized by exchange of infinite towers of KK Higgs and 
resultant amplitude scales linearly with the scattering energy 
$\propto\sqrt{s}$. It has been also found that a tree level partial wave 
unitarity condition is satisfied up to $6.7(5.7)$ TeV for the KK scale 
$m_{KK}=430(500)$ GeV. As mentioned above, simple extensions of the UED, which 
is constructed on the flat five dimensional spacetime, leads to deviations of 
the Higgs coupling. The LHC experiment would check such predictions. How about 
in a warped case with bulk SM Higgs?

Regarding to the deviations of the Higgs coupling from the SM expectation, a 
$SO(5)\times U(1)$ GHU model \cite{Hosotani:2008tx,Hosotani:2008by} lead to 
suppressed coupling for $WWH$, $ZZH$, and Yukawa interactions by a factor 
$\cos\theta_H$ where $\theta_H$ is the Wilson line phase. Since 
$\theta_H=\pi/2$ can be dynamically realized in the model, the couplings for 
the above interactions vanish. As the results, the Higgs becomes stable, and 
thus, it can be a candidate for DM 
\cite{Hosotani:2009jk,Hosotani:2010hx}\footnote{Another candidate for DM in 
$SO(5)\times U(1)_X$ GHU model can be realized by imposing an anti-periodic BC
 for a bulk field presented in \cite{Haba:2009xu}.}. The mass of the Higgs in 
the model is predicted in a region $70\mbox{ GeV}\leq m_H\leq135$ GeV for the 
warp factor $10^5\leq z_L\leq10^{15}$. 

The simplest model with the bulk Higgs on the warped extra-dimension to 
reproduce the SM on the four-dimensional brane is still the {\it bulk SM} 
\cite{Chang:1999nh}. In the work \cite{Chang:1999nh}, two kind of scenarios 
have been proposed. One is that all SM particles live in the bulk and the other
 is that only Higgs is brane field while other SM particles are in the bulk. 
The work has pointed out that the mass of the first KK excitation of the $W$ 
boson should be larger than $9$ TeV, whose constraint comes from the EW 
precision measurements\footnote{In a case that the SM gauge bosons live in the 
bulk while leptons and quarks are on the brane 
\cite{Davoudiasl:1999tf,Pomarol:1999ad}, the KK scalar should be larger than 
$23$ TeV, which is obtained from the EW precision measurements of the leptonic 
width of $Z$, atomic parity violation, and deep inelastic neutrino scattering 
\cite{Davoudiasl:1999tf}.}. This bound is certainly weaken from that of 
\cite{Davoudiasl:1999tf}. However, the work \cite{Chang:1999nh} pointed out 
that the simple bulk SM where all SM field are in the bulk should be discarded.
 The reason is as follows: In the model, the following potential of the 
five-dimensional Higgs field with a negative mass squared is taken,
 \begin{eqnarray}
  \mathcal{V}=\frac{\lambda_5}{2}|\Phi|^4-\mu^2|\Phi|^2,
 \end{eqnarray}
where $\lambda_5$ is a quartic coupling in five dimensions, and the VEV is 
assumed to develop as constant in the bulk. The VEV should generate the bulk 
mass term for the gauge boson, $m_V$. In order to reproduce the gauge boson 
masses and conserve $kz_L^{-1}$ of $\mathcal{O}(\mbox{TeV})$, the magnitude of 
$m_V$ should be around $\mathcal{O}(10^2\mbox{ GeV})$. However, the natural 
value for $m_V$ in the RS background would be the same order as $k$ which is 
about $M_{pl}$. In other word, the gauge hierarchy is not solved at all in this
 simple bulk SM because the smallness of $m_V$ requires small $\mu$, which 
should be around the EW scale. Therefore it was concluded that the brane Higgs 
is the only choice to be able to avoid the above fine-tuning of the Higgs mass 
in the simple extension of the SM to RS background\footnote{Notice that the 
assumption of constant VEV profile in \cite{Chang:1999nh} is difficult to 
realize as shown in Tab. \ref{tab1}.}.
 
Next, let us consider an application of the above consideration to the GW 
mechanism with the bulk SM Higgs. We start with the following question: {\it 
Can the bulk SM Higgs stabilize the size of the warped extra-dimension?} (Here 
we neglect the fine-tuning problem of the bulk Higgs mass.) The answer for the 
question is {\it No}. The reason is as follows. The most essential term in the 
potential of GW mechanism is the second one in \eqref{radion-pot}, which make 
the mechanism work. Appropriate sizes of $v_L$ and $v_0e^{-\epsilon kL}$ makes 
the global minimum in the potential. Then the simple numerical game can give a 
viable example to realize the scenario without extreme fine-tuning among 
parameters as in \eqref{example}. The example seems natural for choice of the 
magnitude of parameters, that is, all parameters are of the order of the Planck
 scale (in the basis where graviton is canonically normalized at the Planck 
brane). This means that since the GW mechanism corresponds to $(D,D)$ type BCs,
 the dynamical scale of the Higgs field on the IR brane is just determined by 
the scale $v_2=v_L$, which is the Planck scale. In other words, the physical 
scale of the Higgs on IR brane such as its mass depends only on the fundamental
 scale. Similar situation has been discussed in the Dirichlet Higgs model 
\cite{Haba:2009pb,Haba:2010xz,Oda:2010te} for a five dimensional flat metric. 
In the model, Higgs mass becomes the compactification scale and the EW symmetry
 breaking occurs at the scale $v_1=v_2=v_L$, which can be taken at the usual EW
 scale in the case. On the other hand, in this application of the Higgs as the 
GW scalar, the exact $(D,D)$ should be taken in order to avoid a strong 
coupling for the boundary quartic interaction for Higgs. This means that there 
is no boundary Higgs potential unlike the original RS setup for the SM sector. 
In a simple bulk SM with the bulk Higgs, if the VEV profile sufficiently 
localizes to the IR brane, the gauge boson masses might be reproduced. Such 
situation can be realized in the $(D,D)$ and $(N,D)$ cases without introducing 
large boundary coupling of the Higgs at IR brane as shown in Tab. \ref{tab1}. 
However, there are non-negligible contributions to the $T$ parameter in such 
kind of models. An introduction of additional symmetry $SU(2)_R\times 
U(1)_{B-L}$ is one of approaches to suppress contributions to the $T$ 
parameter. We focus on the model in the next subsection, and discuss the GW in 
the model.

\subsection{GW mechanism in left-right model with custodial symmetry}

In this section, we study a model with custodial symmetry in warped space. If 
we consider a model with brane Higgs and bulk gauge field, the KK states of 
gauge bosons contribute to the EW observables, which can be understood in terms
 of $S$, $T$, and $U$ parameters \cite{stu1,stu2,stu3}. Such a model is 
severely constrained by the EW precision measurements. The introduction of an 
additional symmetry, $SU(2)_R\times U(1)_{B-L}$, has been proposed to avoid 
such constraints, especially for the $T$ parameter \cite{Agashe:2003zs}.

The model \cite{Agashe:2003zs} starts with the $SU(3)_C\times SU(2)_L\times 
SU(2)_R\times U(1)_{B-L}$ gauge symmetry in the bulk. The additional $SU(2)_R$ 
symmetry is broken to $U(1)_R$ by BCs on gauge fields at the UV brane to 
reproduce the usual EW symmetry while conserving the $SU(2)_R$ at the IR brane.
 Then the remaining $U(1)_R\times U(1)_{B-L}$ is spontaneously broken down to 
$U(1)_Y$ at the UV brane. The bulk action for the gauge and fermion sectors is 
given by
 \begin{eqnarray}
  S=\int d^5x\sqrt{-G}(\mathcal{L}_g+\mathcal{L}_f),
 \end{eqnarray}
where $\mathcal{L}_g$ and $\mathcal{L}_f$ are the Lagrangian for the gauge and 
fermions sectors given as
 \begin{eqnarray}
  \mathcal{L}_g 
   &=& -\frac{1}{4}\sqrt{-G}(\mbox{tr}W_{MN}W^{MN}
                             +\mbox{tr}\tilde{W}_{MN}\tilde{W}^{MN}
                             +\mbox{tr}\tilde{B}_{MN}\tilde{B}^{MN}
                             +\mbox{tr}F_{MN}F^{MN} \nonumber \\
   & & \phantom{-\frac{1}{4}\sqrt{-G}(\mbox{tr}W_{MN}W^{MN}
                                      +\mbox{tr}\tilde{W}_{MN}\tilde{W}^{MN}} 
       +G^{MN}(D_M\Sigma)^\dagger(D_N\Sigma)+\mathcal{V}(\Sigma)), \\
  \mathcal{L}_f
   &=& \sqrt{-G}(i\bar{\Psi}\Gamma^MD_M\Psi-\epsilon(y)c_\Psi\bar{\Psi}\Psi).
 \end{eqnarray}
The $W_{MN}$, $\tilde{W}_{MN}$, $\tilde{B}_{MN}$, and $F_{MN}$ are the field 
strength for $SU(2)_L$, $SU(2)_R$, $U(1)_{B-L}$, and $SU(3)_C$ of gauge groups,
 respectively, and $\Sigma$ is a $SU(2)_R$ triplet Higgs, which breaks 
$SU(2)_R$ to $U(1)_R$ in the bulk at scale below $k$. The $\epsilon(y)$ is the 
sign function and $c_\Psi$ is a parameter which determines the localization of 
the zero mode so that the wave-function profile localizes towards to the UV 
(IR) brane for $c_\Phi>1/2$ $(<1/2)$ \cite{Grossman:1999ra,Gherghetta:2000qt}. 
After the NG mode of $\Sigma$ is eaten by the $SU(2)_R$ gauge field, the action
 for the gauge sector can be rewritten by
 \begin{eqnarray}
  \mathcal{L}_g 
   &=& -\frac{1}{4}\sqrt{-G}(\mbox{tr}W_{MN}W^{MN}
                             +\mbox{tr}\tilde{W}_{MN}\tilde{W}^{MN}
                             +\mbox{tr}\tilde{B}_{MN}\tilde{B}^{MN}
                             +\mbox{tr}F_{MN}F^{MN} \nonumber \\
   & & \phantom{-\frac{1}{4}\sqrt{-G}(\mbox{tr}W_{MN}W^{MN}
                                      +\mbox{tr}\tilde{W}_{MN}\tilde{W}^{MN}
                                      +\mbox{tr}\tilde{B}_{MN}\tilde{B}^{MN}} 
       +\tilde{M}^2|\tilde{W}^\pm|^2).
 \end{eqnarray}
The vanishing $\tilde{M}$ corresponds to the unbroken $SU(2)_R$ in the bulk. 
The UV brane includes fields to break $U(1)_R\times U(1)_{B-L}$ to $U(1)_Y$ and
 the IR brane contains the usual SM Higgs field, which is described by a 
bidoublet under $SU(2)_L\times SU(2)_R$. The assignment for the gauge fields 
under $S^1/Z_2\times Z_2'$ orbifold are given by $(\mbox{UV},\mbox{ IR})=(-,+)$
 for $\tilde{W}_\mu^{1,2}$ and $(+,+)$ for other gauge fields. The 
$U(1)_R\times U(1)_{B-L}\rightarrow U(1)_Y$ breaking occurs through a VEV at 
the UV brane. The gauge fields related to the additional $SU(2)_R$ can be 
written as 
 \begin{eqnarray}
  Z_\mu'\equiv\frac{\tilde{g}_5\tilde{W}_\mu^3-\tilde{g}_5'\tilde{B}_\mu}
                   {\sqrt{\tilde{g}_5^2+\tilde{g}_5'{}^2}}~~~\mbox{ and }~~~
  B_\mu\equiv\frac{\tilde{g}_5'\tilde{W}_\mu^3-\tilde{g}_5\tilde{B}_\mu}
                  {\sqrt{\tilde{g}_5^2+\tilde{g}_5'{}^2}},
 \end{eqnarray} 
where the covariant derivative is defined by $D_M\equiv\partial_M-i(g_5W_M^a\tau_{aL}+\tilde{g}_5\tilde{M}_M^a\tau_{aR}+\tilde{g}_5'\tilde{B}_M\tilde{Y})$ 
with $\tilde{Y}=(B-L)/2$. The hypercharge coupling, $Z_5'$ coupling, and 
$\tilde{B}$-$\tilde{W}^3$ mixing angle are defined by
 \begin{eqnarray}
  g_5'=\frac{\tilde{g}_5'\tilde{g}_5}{\sqrt{\tilde{g}_5^2+\tilde{g}_5'{}^2}},
  ~~~g_{Z'5}=\tilde{g}_5^2+\tilde{g}_5'{}^2,
  ~~~\sin\theta_W'=\frac{\tilde{g}_5'}{g_{Z'5}}.
 \end{eqnarray}
where we can write four dimensional gauge couplings as $g=g_5/\sqrt{L}$, 
$g'=g_5'/\sqrt{L}$, $\tilde{g}=\tilde{g}_5/\sqrt{L}$, 
$\tilde{g}'=\tilde{g}_5'/\sqrt{L}$, and $g_{Z'}=g_{Z'5}/\sqrt{L}$. For the 
fermion sector, the usual right-handed fermions should be promoted to doublets 
under $SU(2)_R$ because the fermions are bulk fields and there is $SU(2)_R$ 
symmetry on the bulk. 

Under the above setup, the EW fit has been discussed in terms of 
four-dimensional effective Lagrangian after integrating out the heavy modes 
\cite{Holdom:1990tc,Golden:1990ig,Georgi:1991ci,Barbieri:1999tm,Csaki:2002gy}. 
The dimension-six operators, 
 \begin{eqnarray}
  \mathcal{L}_6 &=& \frac{1}{16\pi^2v^2}
                    \left[gg'sH^\dagger\tau^aHB^{\mu\nu}W_{a\mu\nu}
                         +(-t)((D^\mu H)^\dagger H)(H^\dagger D_\mu H)
                    \right. \nonumber \\
                & & \phantom{\frac{1}{16\pi^2v^2}[}      
                    \left.
                    +(-ix)\bar{\psi}\gamma^\mu\tau^a\psi(D_\mu H)^\dagger\tau_a
                     H+(-iy)\bar{\psi}\gamma^\mu\psi(D_\mu H)^\dagger H
                    +V\bar{\psi}\psi\bar{\psi}\psi+h.c.\right], \nonumber 
  \label{six} \\
 \end{eqnarray}
are important for the fit, where $x$, $y$, and $V$ generally take as different 
values (couplings) for each fermion. The first and second terms are 
higher-dimensional operators corresponding to the gauge kinetic and mass terms 
for gauge fields, respectively, and the terms in the second line \eqref{six} 
correspond to the fermion sector. When the higher dimensional operators can be 
written as
 \begin{eqnarray}
  x=ag^2~~~\mbox{ and }~~~y=ag'{}^2YY_H,
 \end{eqnarray}
these effects can be translated into oblique parameter by performing the 
following field redefinition,
 \begin{eqnarray}
  && W_3\rightarrow 
      W_3\left(1-g^2\frac{a}{64\pi^2}\right)+Bgg'\frac{a}{64\pi^2},~~~
     W^\pm\rightarrow W^\pm\left(1-g^2\frac{a}{64\pi^2}\right), \\
  && B\rightarrow
      B\left(1-g'{}^2\frac{a}{64\pi^2}\right)+W_3gg'\frac{a}{64\pi^2},
 \end{eqnarray}
where $Y$ and $Y_H$ are hypercharges of each fermion and Higgs, respectively. 
Then, the parameters $s$ and $t$, and this redefinition give
 \begin{eqnarray}
  S=\frac{s}{2\pi}+\frac{a}{2\pi},~~~\mbox{ and }~~~
  T=\frac{t}{8\pi e^2}+\frac{ag'{}^2}{8\pi e^2}.
 \end{eqnarray}
Finally, the $S$ and $T$ parameters in this model can be estimated as 
\cite{Agashe:2003zs}
 \begin{eqnarray}
  S\simeq2\pi\left(\frac{v_{\text{EW}}}{k}z_L\right)^2~~~\mbox{ and }~~~
  T\simeq\frac{\pi}{2}\frac{\tilde{g}^2}{e^2}
         \left(\frac{v_{\text{EW}}}{k}z_L\right)^2kL\frac{\tilde{M}^2}{4k^2}. 
 \end{eqnarray} 
Notice that the main contribution to $T$ is proportional to the bulk mass of 
$SU(2)_R$ gauge boson. Therefore, if the bulk $SU(2)_R$ is unbroken, the main 
contribution becomes the next leading order of $\mathcal{O}((kL)^0)$.

The above model with the bulk custodial symmetry is one of fascinating models 
to be extended from the SM to five-dimensional model on warped space. This can 
pass severe EW precision tests due to the custodial protection and have some 
predictions for collider signatures because of a lower bound of the KK scale 
around a few TeV. Our next task is to consider a simplification of warped 
extra-dimensional models. Towards this purpose we start with the question: {\it
 Can the triplet Higgs under $SU(2)_R$ stabilize the radius of extra 
dimension?} The answer for the question is {\it Yes}. The essential points for 
the GW mechanism are \eqref{example}, namely the VEVs around the 
five-dimensional fundamental scale of order the Planck one and slightly smaller
 bulk mass of the triplet Higgs than the AdS curvature scale. On the other 
hand, in order to suppress the contribution from the KK sector to the $T$ 
parameter in the framework of the model with the custodial symmetry, unbroken 
$SU(2)_R$ in the bulk is favored. Of course, broken $SU(2)_R$ with 
$\tilde{M}\ll k$ is also possible as mentioned above. The point for working GW 
mechanism and that of custodial protection can be completely compatible with 
each other. Therefore, we can achieve a {\it radius stabilization by the 
$SU(2)_R$ triplet Higgs}. In this scenario, the $SU(2)_R$ is broken at both 
boundaries by appropriate BCs which should be either Neumann BCs with large 
quartic Higgs coupling at boundaries or $(D,D)$ BCs to stabilize the radius as 
in the original GW mechanism. And we can always take the bulk potential 
$\mathcal{V}(\Sigma)$ of the triplet Higgs as favored for the custodial 
protection. In this direction, an additional bulk scalar is not needed. 
Therefore, we conclude that this scenario is one of the simplest extensions of 
the SM to five-dimensional model on the warped space to realize radius 
stabilization and easily pass the EW precision tests without any fine-tunings.

We have shown that the GW radius stabilization can be achieved by the $SU(2)_R$
 triplet Higgs in the $(D,D)$ case (or corresponding replacement to Neumann BC 
with large boundary coupling) as one of the simplest extensions of the SM to 
five-dimensional model on the warped space. The introduction of the triplet 
Higgs is one of options discussed in the model \cite{Agashe:2003zs}. The sole 
role of the field is to spontaneously break $SU(2)_R$ to $U(1)_R$ at a mass 
scale below curvature scale. However, it is not necessary for the protection of
 $T$ parameter to introduce the field, rather, an option without the triplet 
(unbroken $SU(2)_R$) is more favored for the protection. Therefore, the model 
without the triplet is still simple where the radius stabilization is realized 
by the conventional GW mechanism (with gauge singlet bulk scalar). Finally, we 
comment on other possibility of extension of the SM. In a left-right model with
 custodial symmetry \cite{Agashe:2003zs}, the SM Higgs exactly localizes at the
 IR brane. It might be still possible that this Higgs becomes bulk field if the
 wave function sufficiently localizes towards the IR brane so that the gauge 
boson masses are reproduced. When the $(D,D)$ or $(N,D)$ with large boundary 
boundary coupling at UV brane are taken, the bulk Higgs could stabilize the 
radius of extra-dimension. The boundary quartic coupling at IR brane should not
 be large because the Higgs corresponds to the SM one in this case. In such 
models, one would have to discuss generating Yukawa hierarchies and flavor 
changing neutral currents because an overlap among the Higgs and fermions near
 the UV brane are not suppressed. Such considerations would be worth studying 
further.

\section{Summary}

We have studied implications of generalized non-zero Dirichlet BC along with 
the ordinary Neumann one on a bulk scalar in the RS warped compactification. 
First we have shown profiles of VEV of the scalar under the general BCs. These 
BCs are described by combinations of the Neumann and Dirichlet types as, 
$(\mbox{UV},\mbox{IR})=(D,D)$, $(D,N)$, $(N,D)$, and $(N,N)$. It has been 
clarified that the VEV profile localizes toward to the IR brane in the $(D,D)$ 
and $(N,D)$ BCs while it localizes toward to the UV brane in the $(D,N)$ case.
 And we have also shown that there is not a non-zero solution of EOM in the 
$(N,N)$ case.

We have also investigated GW mechanism in several setups with the general 
boundary conditions of the bulk scalar field. We have shown that the GW 
mechanism can work under non-zero Dirichlet BCs with appropriate size of VEVs. 
(i) First we have considered the application: the bulk SM Higgs as a bulk 
scalar for the GW mechanism. We have also reviewed related topics: In this 
application, the $(D,D)$ BCs should be taken in order to avoid a strong 
coupling for the boundary quartic interaction for the Higgs while realizing the
 GW mechanism. Furthermore, the bulk SM where all SM field live in the bulk 
cannot still solve the hierarchy problem because of the required bulk mass of 
the order of the EW scale to reproduce the SM gauge boson masses. This 
difficulty cannot be avoided under all combinations of BCs formulated in the 
above even if the brane localized Higgs potential are introduced. Therefore, we 
conclude that the bulk SM Higgs cannot be a GW stabilizer unless we allow 
unnaturally small bulk mass compared to the fundamental scale. (ii) We have also
 discussed an application of the triplet Higgs under additional $SU(2)_R$ 
symmetry to a bulk scalar in the GW mechanism. In this scenario, all the 
requirements to realize the GW mechanism and custodial protection for $T$ 
parameter in $SU(3)_C\times SU(2)_L\times SU(2)_R\times U(1)_{B-L}$ model are 
completely compatible with each other. Therefore, we conclude that the radius 
can be stabilized by the $SU(2)_R$ triplet Higgs. 

\subsection*{Acknowledgments}

We would like to thank T. Yamashita for very helpful comments. This work is 
partially supported by Scientific Grant by Ministry of Education and Science, 
Nos.\ 20540272, 20039006, 20025004, 20244028, and 19740171. The work of RT is 
supported by the DFG-SFB TR 27. 
 
\appendix
\section{Case with brane localized potentials}
\label{A0}

\subsection{Action and BCs}

In this appendix, we give formulation and VEV profile in a case with brane 
localized scalar potentials. We should start with the following action of the 
bulk scalar in behalf of \eqref{action1}, 
 \begin{eqnarray}
  S=\int d^4x\int_0^Ldye^{-4\sigma}[-e^{2\sigma}|\partial_\mu\Phi|^2
    -|\partial_y\Phi|^2
    -\mathcal{V}-\delta(y)V_0-\delta(y-L)V_L].
  \label{action1-app}
 \end{eqnarray}
The variation of the action is given by
 \begin{eqnarray}
  \delta S &=& \int d^4x\int_0^Ldye^{-4\sigma}
                     \bigg[\delta\Phi_X\left(\mathcal{P}\Phi_X
                     -\frac{\partial\mathcal{V}}{\partial\Phi_X}\right) 
               \nonumber \\
           & & 
                     +\delta(y)\delta\Phi_X\left(+\partial_y\Phi_X
                     -\frac{\partial V_0}{\partial\Phi_X}
                     \right)
                     +\delta(y-L)\delta\Phi_X\left(-\partial_y\Phi_X
                     -\frac{\partial V_L}{\partial\Phi_X}
                     \right)\bigg],
 \end{eqnarray}
and thus the Neumann BC should be modified from \eqref{Neumann} to
 \begin{eqnarray}
  \left.\pm\partial_y\Phi_X
  -\frac{\partial V_\eta}{\partial\Phi_X}
  \right|_{y=\eta}=0, 
 \end{eqnarray}
while the Dirichlet one is the same as \eqref{Dirichlet} even if the brane 
localized potentials are introduced. In this Appendix, we take the brane 
localized potentials as
 \begin{eqnarray}
  V_\eta=\lambda_\eta\left(|\Phi|^2-\frac{v_\eta^2}{2}\right)^2
       =\frac{\lambda_\eta}{4}(\Phi_R^2+\Phi_I^2-v_\eta^2)^2. \label{brane-pot}
 \end{eqnarray}

The free action for the physical Higgs and NG can be written as
 \begin{eqnarray}
  S &=& \int d^4x\int_0^Ldye^{-4\sigma}
        \Bigg[-\mathcal{V}
              -\frac{e^{2\sigma}}{2}(\partial_\mu\Phi_R+\partial_\mu\phi)^2
              -\frac{e^{2\sigma}}{2}(\partial_\mu\Phi_I+\partial_\mu\chi)^2
        \nonumber \\ 
    & & -\frac{1}{2}(\partial_y\Phi_R+\partial_y\phi)^2
        -\frac{1}{2}(\partial_y\Phi_I+\partial_y\chi)^2
        -\frac{\partial\mathcal{V}}{\partial\Phi_R}^c\phi
        -\frac{\partial\mathcal{V}}{\partial\Phi_I}^c\chi
        -\frac{1}{2}\frac{\partial^2\mathcal{V}}{\partial\Phi_R^2}^c\phi^2
        -\frac{1}{2}\frac{\partial^2\mathcal{V}}{\partial\Phi_I^2}^c\chi^2
        \nonumber \\
    & &  -\delta(y)\Bigg(V_0+\frac{\partial V_0}{\partial\Phi_R}^c\phi
                         +\frac{\partial V_0}{\partial\Phi_I}^c\chi
                         +\frac{1}{2}\frac{\partial^2V_0}{\partial\Phi_R^2}^c
                          \phi^2
                         +\frac{1}{2}\frac{\partial^2V_0}{\partial\Phi_I^2}^c
                          \chi^2\Bigg) \nonumber \\
    & & -\delta(y-L)\Bigg(V_L+\frac{\partial V_L}{\partial\Phi_R}^c\phi
                          +\frac{\partial V_L}{\partial\Phi_I}^c\chi
                          +\frac{1}{2}\frac{\partial^2V_L}{\partial\Phi_R^2}^c
                           \phi^2
                          +\frac{1}{2}\frac{\partial^2V_L}{\partial\Phi_I^2}^c
                           \chi^2\Bigg)\Bigg].
 \end{eqnarray}
The partial integrals for 
$-e^{-2\sigma}(\partial_\mu\Phi_R)(\partial_\mu\phi)$, 
$-e^{-2\sigma}(\partial_\mu\Phi_I)(\partial_\mu\chi)$, 
$-e^{-4\sigma}(\partial_y\Phi_R)(\partial_y\phi)$ and 
$-e^{-4\sigma}(\partial_y\Phi_I)(\partial_y\chi)$ in the integrand give
 \begin{eqnarray}
  S &=& \int d^4x\int_0^Ldye^{-4\sigma}
        \Bigg[-\mathcal{V}
              -\frac{e^{2\sigma}}{2}
               \{(\partial_\mu\Phi_R)^2-2\phi\Box\Phi_R+(\partial_\mu\phi)^2\}
        \nonumber \\
    & & -\frac{e^{2\sigma}}{2}
         \{(\partial_\mu\Phi_I)^2-2\chi\Box\Phi_I+(\partial_\mu\chi)^2\}
        -\frac{1}{2}\{(\partial_y\Phi_R)^2
                      -2e^{4\sigma}\phi\partial_y(e^{-4\sigma}(\partial_y\Phi_R))
                      +(\partial_y\phi)^2\} \nonumber \\
    & & -\frac{1}{2}\{(\partial_y\Phi_I)^2
                      -2e^{4\sigma}\chi\partial_y(e^{-4\sigma}(\partial_y\Phi_I))
                      +(\partial_y\chi)^2\}
        -\frac{\partial\mathcal{V}}{\partial\Phi_R}^c\phi
        -\frac{\partial\mathcal{V}}{\partial\Phi_I}^c\chi \nonumber \\
    & & -\frac{1}{2}\frac{\partial^2\mathcal{V}}{\partial\Phi_R^2}^c\phi^2
        -\frac{1}{2}\frac{\partial^2\mathcal{V}}{\partial\Phi_I^2}^c\chi^2
        \nonumber \\
    & & -\delta(y)\Bigg(V_0-\phi\partial_y\Phi_R-\chi\partial_y\Phi_I
                        +\frac{\partial V_0}{\partial\Phi_R}^c\phi
                        +\frac{\partial V_0}{\partial\Phi_I}^c\chi
                        +\frac{1}{2}\frac{\partial^2V_0}{\partial\Phi_R^2}^c
                         \phi^2
                        +\frac{1}{2}\frac{\partial^2V_0}{\partial\Phi_I^2}^c
                         \chi^2\Bigg) \nonumber \\
    & & -\delta(y-L)\Bigg(V_L+\phi\partial_y\Phi_R+\chi\partial_y\Phi_I
                          +\frac{\partial V_L}{\partial\Phi_R}^c\phi
                          +\frac{\partial V_L}{\partial\Phi_I}^c\chi
                          +\frac{1}{2}\frac{\partial^2V_L}{\partial\Phi_R^2}^c
                           \phi^2
                          +\frac{1}{2}\frac{\partial^2V_L}{\partial\Phi_I^2}^c
                           \chi^2\Bigg)\Bigg]. \nonumber \\
 \end{eqnarray}
Here, note that the terms depending only on $\Phi_X$, $V_\eta$, and 
$\mathcal{V}$ are vanishing due to the EOM and Neumann BCs. Therefore, we can 
obtain the following action,
 \begin{eqnarray}
  S &=& \int d^4x\int_0^Ldye^{-4\sigma}
        \Bigg[-\frac{e^{2\sigma}}{2}\{-2\phi\Box\Phi_R+(\partial_\mu\phi)^2\}
              -\frac{e^{2\sigma}}{2}\{-2\chi\Box\Phi_I+(\partial_\mu\chi)^2\}
        \nonumber \\
    & & -\frac{1}{2}\{-2e^{4\sigma}\phi\partial_y(e^{-4\sigma}(\partial_y\Phi_R))
                      +(\partial_y\phi)^2\}
        -\frac{1}{2}\{-2e^{4\sigma}\chi\partial_y(e^{-4\sigma}(\partial_y\Phi_I))
                      +(\partial_y\chi)^2\} \nonumber \\ 
    & & -\frac{\partial\mathcal{V}}{\partial\Phi_R}^c\phi
        -\frac{\partial\mathcal{V}}{\partial\Phi_I}^c\chi
        -\frac{1}{2}\frac{\partial^2\mathcal{V}}{\partial\Phi_R^2}^c\phi^2
        -\frac{1}{2}\frac{\partial^2\mathcal{V}}{\partial\Phi_I^2}^c\chi^2
        \nonumber \\
    & & -\delta(y)\Bigg(-\phi\partial_y\Phi_R-\chi\partial_y\Phi_I
                        +\frac{\partial V_0}{\partial\Phi_R}^c\phi
                        +\frac{\partial V_0}{\partial\Phi_I}^c\chi
                        +\frac{1}{2}\frac{\partial^2V_0}{\partial\Phi_R^2}^c
                         \phi^2
                        +\frac{1}{2}\frac{\partial^2V_0}{\partial\Phi_I^2}^c
                         \chi^2\Bigg) \nonumber \\
    & & -\delta(y-L)\Bigg(+\phi\partial_y\Phi_R+\chi\partial_y\Phi_I
                          +\frac{\partial V_L}{\partial\Phi_R}^c\phi
                          +\frac{\partial V_L}{\partial\Phi_I}^c\chi
                          +\frac{1}{2}\frac{\partial^2V_L}{\partial\Phi_R^2}^c
                           \phi^2
                          +\frac{1}{2}\frac{\partial^2V_L}{\partial\Phi_I^2}^c
                           \chi^2\Bigg)\Bigg]. \nonumber \\
 \end{eqnarray}
We also notice that the linear terms of $\phi$ and $\chi$ vanishes because of 
the EOM and Neumann BCs for the $\Phi_X$ fields as
 \begin{eqnarray}
  S &=& \int d^4x\int_0^Ldye^{-4\sigma}
        \Bigg[-\frac{e^{2\sigma}}{2}(\partial_\mu\phi)^2
              -\frac{e^{2\sigma}}{2}(\partial_\mu\chi)^2
              -\frac{1}{2}(\partial_y\phi)^2-\frac{1}{2}(\partial_y\chi)^2
        \nonumber \\
    & & -\frac{1}{2}\frac{\partial^2\mathcal{V}}{\partial\Phi_R^2}^c\phi^2
        -\frac{1}{2}\frac{\partial^2\mathcal{V}}{\partial\Phi_I^2}^c\chi^2
        \nonumber \\
    & & -\delta(y)\Bigg(\frac{1}{2}
                        \frac{\partial^2V_0}{\partial\Phi_R^2}^c\phi^2
                        +\frac{1}{2}\frac{\partial^2V_0}{\partial\Phi_I^2}^c
                         \chi^2\Bigg) 
        -\delta(y-L)\Bigg(\frac{1}{2}\frac{\partial^2V_L}{\partial\Phi_R^2}^c
                          \phi^2
                          +\frac{1}{2}\frac{\partial^2V_L}{\partial\Phi_I^2}^c
                           \chi^2\Bigg)\Bigg]. 
 \end{eqnarray} 
The partial integrals for each kinetic term make the action
 \begin{eqnarray}
  S &=& \int d^4x\int_0^Ldye^{-4\sigma}
        \Bigg[\frac{1}{2}\phi
              \Bigg(e^{2\sigma}\Box
                    +e^{4\sigma}\partial_ye^{-4\sigma}\partial_y
                    -\frac{\partial^2\mathcal{V}}{\partial\Phi_R^2}^c\Bigg)\phi
        \nonumber \\
    & & +\frac{1}{2}\chi
         \Bigg(e^{2\sigma}\Box
               +e^{4\sigma}\partial_ye^{-4\sigma}\partial_y
               -\frac{\partial^2\mathcal{V}}{\partial\Phi_I^2}^c\Bigg)\chi
        \nonumber \\
    & & -\frac{\delta(y)}{2}
         \Bigg(-\phi\partial_y\phi-\chi\partial_y\chi
               +\frac{\partial^2V_0}{\partial\Phi_R^2}^c\phi^2
               +\frac{\partial^2V_0}{\partial\Phi_I^2}^c\chi^2\Bigg) 
        \nonumber \\
    & & -\frac{\delta(y-L)}{2}
         \Bigg(+\phi\partial_y\phi+\chi\partial_y\chi
               +\frac{\partial^2V_L}{\partial\Phi_R^2}^c\phi^2
               +\frac{\partial^2V_L}{\partial\Phi_I^2}^c\chi^2\Bigg)\Bigg].
 \end{eqnarray}
It is seen that the Neumann BCs are modified from \eqref{Neumann-quantum} as
 \begin{eqnarray}
  \left.\left(\pm\partial_y-\frac{\partial^2V_\eta}{\partial\Phi_R^2}^c\right)
  f_n(y)\right|_{y=\eta}=0.
 \end{eqnarray}
 
\subsection{VEV profiles}

We discuss the VEV profiles for the $(D,N)$ and $(N,D)$ BCs in the case that 
the boundary quartic coupling is finite.

\subsubsection{{\boldmath$(D,N)$} case}

The BCs are given by
 \begin{eqnarray}
  v(1) = v_1,~~~ 
  \partial_zv(z)|_{z=z_L}
  +\left.\frac{\partial V_L}{\partial\Phi}^c\right|_{z=z_L} = 0.
 \end{eqnarray}
They are written down as
 \begin{eqnarray}
  &&A+B=v_1, \label{BC-DN-0} \\
  &&k[A(\nu+2)z_L^{\nu+2}-B(\nu-2)z_L^{-(\nu-2)}] \nonumber \\
  &&+\lambda_L(Az_L^{\nu+2}+Bz_L^{-(\nu-2)})
     [(Az_L^{\nu+2}+Bz_L^{-(\nu-2)})^2-v_L^2]=0. \label{BC-DN}
 \end{eqnarray}
One must note that when the coupling in the boundary potential, $\lambda_\eta$,
 becomes infinite, the Neumann BCs turn to the Dirichlet ones, 
$v(y)|_{y=\eta}\rightarrow v_\eta$. Similar situation has been discussed in the
 GW mechanism \cite{Goldberger:1999uk}.

When the boundary quartic coupling is finite, the numerical calculation 
indicates $A\ll B\simeq z_L^{\nu-2}v_L$ as a solution of \eqref{BC-DN-0} and 
\eqref{BC-DN}. Then the VEV profile can be approximated by
 \begin{eqnarray}
  v(z)\simeq\left(v_1-z_L^{\nu-2}\sqrt{v_L^2+\frac{k(\nu-2)z_L^{-(\nu-2)}}{\lambda_L}}\right)z^{\nu+2}+\sqrt{v_L^2+\frac{k(\nu-2)z_L^{-(\nu-2)}}{\lambda_L}}\left(\frac{z_L}{z}\right)^{\nu-2}.
 \end{eqnarray}
A typical behavior of VEV profile in a case that the boundary coupling is 
finite is shown in the left figure of Fig. \ref{fig0}.

\subsubsection{{\boldmath$(N,D)$ case}}

The BCs are
 \begin{eqnarray}
  \partial_z v(z)|_{z=1}+\left.\frac{\partial V_0}{\partial\Phi}^c\right|_{z=1}
   = 0,~~~
  v(z_L) = v_2.
 \end{eqnarray}
They are written down as
 \begin{eqnarray}
  k[A(\nu+2)-B(\nu-2)]-\lambda_0(A+B)[(A+B)^2-v_0^2] &=& 0, \\
  Az_L^{\nu+2}+Bz_L^{-(\nu-2)} &=& v_2. 
 \end{eqnarray}
Numerical calculation indicates $A\sim B\sim v_2/z_L^{\nu+2}$ and the VEV 
profile is approximated by
 \begin{eqnarray}
  v(z)\simeq\left[v_2-\frac{(k+\lambda_0v_0v_2)v_0}
                           {2k\nu-\lambda_0v_0^2(z_L^{2\nu}-1)}\right]
            \left(\frac{z}{z_L}\right)^{\nu+2}
            +\frac{(k+\lambda_0v_0v_2)v_0}{2k\nu-\lambda_0v_0^2(z_L^{2\nu}-1)}
             \left(\frac{z}{z_L}\right)^{-(\nu-2)}.
 \end{eqnarray}
The typical VEV profile is shown in the lest figure of Fig. \ref{fig0}.

\section{Gauge sector}
\label{B} 

In this Appendix, we write down interactions of SM gauge field with the bulk SM
 Higgs to show the dependence of the gauge boson mass on the VEV profile. Then 
a deconstruction method is also presented, which gives profile of gauge field. 

\subsection{Interactions of gauge field with the bulk SM Higgs}

First, we write down interactions of gauge field with the bulk SM Higgs. If the
 Higgs sector are also on the bulk, the Higgs kinetic term is
 \begin{eqnarray}
  S_{kin}&=&\int d^4x\int_0^Ldy\sqrt{-G}[-G^{MN}(D_M\Phi)^\dagger(D_N\Phi)],
 \end{eqnarray}
where
 \begin{eqnarray}
  \Phi&\equiv&\left(
                       \begin{array}{c}
                        \varphi^+(x,y) \\
                        \frac{v(y)+H(x,y)+i\chi(x,y)}{\sqrt{2}}
                       \end{array}
                     \right), \\
  D_M&\equiv&\partial_M+ig_5W_M^aT^a+ig_5'B_MY\\
               &=&\partial_M+i\frac{g_5}{2}
                     \left(
                       \begin{array}{cc}
                        W_M^3            & W_M^1-iW_M^2 \\
                        W_M^1+iW_M^2 & -W_M^3
                       \end{array}
                     \right)+i\frac{g_5'}{2}
                     \left(
                       \begin{array}{cc}
                        B_M & 0 \\
                        0     & B_M
                       \end{array}
                     \right)\\
              &=&\partial_M+i\frac{g_5}{\sqrt{2}}
                     \left(
                       \begin{array}{cc}
                        0        & W_M^+ \\
                        W_M^- & 0
                       \end{array}
                     \right)+\frac{i}{2}
                     \left(
                       \begin{array}{cc}
                        g_5W_M^3+g_5'B_M & 0 \\
                        0     & -g_5W_M^3+g_5'B_M
                       \end{array}
                     \right), \label{cd}
 \end{eqnarray}
and $g_5$ and $g_5'$ are the gauge couplings in five dimension. Here we can 
write the Weinberg angle and the gauge couplings as
 \begin{eqnarray}
  &\sin\theta_W\equiv s_W\equiv\frac{g_5'}{\sqrt{g_5^2+g_5'{}^2}}
                               =\frac{g'}{\sqrt{g^2+g'{}^2}},~~~
  \cos\theta_W\equiv c_W\equiv\frac{g_5}{\sqrt{g_5^2+g_5'{}^2}}
                               =\frac{g}{\sqrt{g^2+g'{}^2}},&\\
  &e\equiv \frac{g_5g_5'}{\sqrt{g_5^2+g_5'{}^2}}
    =\frac{gg'}{\sqrt{g^2+g'{}^2}},~~~
   g_5\equiv g\sqrt{L},~~~
   g_5'\equiv g'\sqrt{L}.&
 \end{eqnarray}
Then the covariant derivative \eqref{cd} can be rewritten by
 \begin{eqnarray}
  D_M&=&\partial_M+i\frac{g_5}{\sqrt{2}}
                     \left(
                       \begin{array}{cc}
                        0        & W_M^+ \\
                        W_M^- & 0
                       \end{array}
                     \right)\nonumber\\
        &&+\frac{i}{2}\sqrt{g_5^2+g_5'{}^2}
                     \left(
                       \begin{array}{cc}
                        c_WW_M^3+s_WB_M & 0 \\
                        0     & -c_WW_M^3+s_WB_M
                       \end{array}
                     \right).
 \end{eqnarray}
Further, after defining
 \begin{eqnarray}
  \left(
   \begin{array}{c}
    Z_M \\
    A_M
   \end{array}
  \right)\equiv\left(
                     \begin{array}{cc}
                      c_W & -s_W \\
                      s_W & c_W
                     \end{array}
                    \right)\left(
                              \begin{array}{c}
                               W_M^3 \\
                               B_M
                             \end{array}
                            \right),
 \end{eqnarray}
The covariant derivative becomes
 \begin{eqnarray}
  D_M=\partial_M+i\frac{g_5}{\sqrt{2}}
                     \left(
                       \begin{array}{cc}
                        0     & W_M^+ \\
                        W_M^- & 0
                       \end{array}
                     \right)+i
                     \left(
                       \begin{array}{cc}
                        \frac{g_5^2-g_5'{}^2}{2\sqrt{g_5^2+g_5'{}^2}}Z_M+eA_W & 0 \\
                        0     & -\frac{\sqrt{g_5^2+g_5'{}^2}}{2}Z_M
                       \end{array}
                     \right).
 \end{eqnarray}
Here, we give useful relations among the Weinberg angle and gauge couplings,
  \begin{eqnarray}
  &&\frac{g_5^2-g_5'{}^2}{2\sqrt{g_5^2+g_5'{}^2}}
  =\frac{\sqrt{g_5^2+g_5'{}^2}(c_W^2-s_W^2)}{2}
  =\frac{c_W^2-s_W^2}{2s_Wc_W}e=\frac{1}{\tan2\theta_W},\\
  &&\sqrt{g_5^2+g_5'{}^2}=\frac{e}{c_Ws_W},~~~g_5=\frac{e}{s_W}.
 \end{eqnarray}
Finally, we obtain
 \begin{eqnarray}
  D_M&=&\partial_M+\frac{i}{\sqrt{2}}\frac{e}{s_W}
                     \left(
                       \begin{array}{cc}
                        0     & W_M^+ \\
                        W_M^- & 0
                       \end{array}
                     \right)+ie
                     \left(
                       \begin{array}{cc}
                        \frac{1}{\tan2\theta_W}Z_M+A_W & 0 \\
                        0     & -\frac{1}{\sin2\theta_W}Z_M
                       \end{array}
                     \right),
 \end{eqnarray}
and thus,
 \begin{eqnarray}
  &&(D_M\Phi)^\dagger(D_N\Phi)\nonumber\\
  &&=\left(\partial_M\varphi^--\frac{ie}{2s_W}W_M^-(v(y)+H-i\chi)              
           -ie\left(\frac{1}{\tan2\theta_W}Z_M+A_M\right)\varphi^-\right)
    \nonumber\\
  &&\phantom{=}\times\left(\partial_N\varphi^++\frac{ie}{2s_W}W_N^-(v(y)+H
               +i\chi)+ie\left(\frac{1}{\tan2\theta_W}Z_N+A_N\right)
               \varphi^+\right)\nonumber\\
  &&\phantom{=}+\frac{1}{2}
     \left(\partial_Mv(y)+\partial_MH-i\partial_M\chi
     -\frac{ie}{s_W}W_M^+\varphi^-+\frac{ie}{\sin2\theta_W}Z_M(v(y)+H-i\chi)
     \right)\nonumber\\
  &&\phantom{=}\times\frac{1}{2}
     \left(\partial_Nv(y)+\partial_NH+i\partial_N\chi
           +\frac{ie}{s_W}W_N^-\varphi^+
           -\frac{ie}{\sin2\theta_W}Z_N(v(y)+H+i\chi)\right).
 \end{eqnarray}
The quadratic, cubic, and quartic terms are written down by
 \begin{eqnarray}
  &&\left.(D_M\Phi)^\dagger(D_N\Phi)\right|_{\mbox{{\scriptsize quadratic}}}
      \nonumber\\
  &&=(\partial_M\varphi^-)(\partial_N\varphi^+)
      +\frac{e}{2s_W}\left[iv(y)(\partial_M\varphi^-)W_N^+
      -iv(y)(\partial_N\varphi^+)W_M^-\right]\nonumber\\
  &&\phantom{=}+\frac{e^2}{4s_W}v(y)^2W_N^+W_M^-
    +\frac{1}{2}(\partial_Mv(y))(\partial_Nv(y))
    +\frac{1}{2}(\partial_Nv(y))(\partial_MH)\nonumber\\
  &&\phantom{=}+\frac{1}{2}(\partial_Mv(y))(\partial_NH)
    +\frac{1}{2}\left[(\partial_MH)(\partial_NH)
    +(\partial_M\chi)(\partial_N\chi)\right]\nonumber\\
  & &\phantom{=}+\frac{e}{s_W}[i(\partial_Mv(y))W_N^-\varphi^+
     -i(\partial_Nv(y))W_M^+\varphi^-]\nonumber\\
  & &\phantom{=}+\frac{e}{2\sin2\theta_W}
     \left[iZ_M\left\{(\partial_Nv(y))v(y)
     +(\partial_Nv(y))(H-i\chi)+(\partial_NH+i\partial_N\chi)v(y)\right\}
     \right.\nonumber\\
  & &\phantom{=+\frac{e}{2\sin2\theta_W}[}\left.
     -iZ_N\left\{(\partial_Mv(y))v(y)
     +(\partial_Mv(y))(H+i\chi)+(\partial_MH-i\partial_M\chi)v(y)\right\}
     \right] \nonumber\\
  & &\phantom{=}+\frac{e^2}{2(\sin2\theta_W)^2}v(y)^2Z_MZ_N,\label{gauge-quad}
 \end{eqnarray}
 \begin{eqnarray}
  &&\left.(D_M\Phi)^\dagger(D_N\Phi)\right|_{\mbox{{\scriptsize cubic}}}
      \nonumber\\
  &&=\frac{e}{2s_W}\left[iW_N(\partial_M\varphi^-)(H+i\chi)
     -iW_M(\partial_N\varphi^+)(H-i\chi)\right]\nonumber\\
  &&\phantom{=}+e\left[i(\partial_M\varphi^-)\varphi^+
     \left(\frac{Z_N}{\tan2\theta_W}+A_N\right)-i(\partial_N\varphi^+)\varphi^-
     \left(\frac{Z_M}{\tan2\theta_W}+A_M\right)\right] \nonumber\\
  &&\phantom{=}+\frac{e^2}{2s_W^2}v(y)W_N^+W_M^-H\nonumber\\
  &&\phantom{=}
     +\frac{e^2}{2s_W}\left[v(y)W_N^+\left(\frac{Z_M}{\tan2\theta_W}
     +A_M\right)\varphi^-+v(y)W_M^-\left(\frac{Z_N}{\tan2\theta_W}
     +A_N\right)\varphi^+\right] \nonumber\\
  &&\phantom{=}
    +\frac{e}{2s_W}\left[i(\partial_MH-i\partial_M\chi)W_N^-\varphi^+
    -i(\partial_NH+i\partial_N\chi)W_M^+\varphi^-\right] \nonumber\\
  &&\phantom{=}
    +\frac{e}{2\sin2\theta_W}\left[iZ_M(\partial_NH+i\partial_N\chi)(H-i\chi)
    -iZ_N(\partial_MH-i\partial_M\chi)(H+i\chi)\right]\nonumber\\
  &&\phantom{=}-\frac{e^2}{2s_W\sin2\theta_W}
    \left[W_N^-\varphi^+Z_Nv(y)+W_M^+\varphi^-Z_Nv(y)\right]      
    +\frac{e^2}{(\sin2\theta_W)^2}v(y)HZ_MZ_N,
 \end{eqnarray}
 \begin{eqnarray}
  &&\left.|(D_M\Phi)^\dagger(D_N\Phi)\right|_{\mbox{{\scriptsize quartic}}}
    \nonumber\\
  &&=\frac{e^2}{4s_W^2}W_N^+W_M^-(H^2+\chi^2)
     +\frac{e^2}{2(\sin\theta_W)^2}Z_NZ_M(H^2+\chi^2)\nonumber\\
  &&\phantom{=}
    +\frac{e^2}{2s_W}
    \left[W_N^+(H+i\chi)\left(\frac{Z_M}{\tan2\theta_W}+A_M\right)\varphi^-
    +W_M^+(H-i\chi)\left(\frac{Z_N}{\tan2\theta_W}+A_N\right)\varphi^+
    \right]\nonumber\\
  &&\phantom{=}+e^2\left(\frac{Z_M}{\tan2\theta_W}+A_M\right)
    \left(\frac{Z_N}{\tan2\theta_W}+A_N\right)\varphi^+\varphi^-
    +\frac{e^2}{2s_W^2}W_N^+W_M^-\varphi^+\varphi^-\nonumber\\
  &&\phantom{=}-\frac{e^2}{2s_W\sin2\theta_W}
    \left[W_N^-\varphi^+Z_M(H-i\chi)+W_M^-\varphi^-Z_N(H+i\chi)\right],
 \end{eqnarray}
where we take $W_y(x,y)^\pm=Z_y(x,y)=0$. We find from \eqref{gauge-quad} that 
the gauge boson masses in \eqref{gauge-mass} is obtained.

\subsection{Deconstruction}

{}For ones who are interested in obtaining gauge field profile, which is not 
directly related with the main discussion of this paper, we show a 
deconstruction method in Abelian case with flat five-dimensional setup. An 
extension to a warped case is straightforward. 

The gauge kinetic term in Abelian case can be written as 
 \begin{eqnarray}
  \int_0^Ldy\mathcal{L}
  &=&-\frac{1}{4}\int_0^Ldy(\partial_MA_N-\partial_NA_M)^2\\
  &=&-\frac{1}{4}\int_0^Ldy\left[(\partial_\mu A_\nu-\partial_\nu A_\mu)^2
     +(\partial_\mu A_y)^2+(\partial_yA_\mu)^2
     +A_y\partial_y\partial_\mu A^\mu\right] \nonumber\\
   & &+\frac{1}{4}[A_y\partial_\mu A^\mu]_0^L,\label{gkt}
 \end{eqnarray}
where we operated an partial integral in the second line. If we take the 
Lorentz gauge, $\partial_\mu A^\mu=0$, as a gauge fixing, \eqref{gkt} turns to
 \begin{eqnarray}
  \int_0^Ldy\mathcal{L}
  =-\frac{1}{4}\int_0^Ldy(\partial_\mu A_\nu-\partial_\nu A_\mu)^2
    -\frac{1}{4}\int_0^Ldy[(\partial_\mu A_y)^2+(\partial_yA_\mu)^2].
  \label{gkt1}
 \end{eqnarray}
After latticizing the five-dimensional coordinate as
 \begin{eqnarray}  
  y_n\equiv na,~~~A_{\mu,n}\equiv A_\mu(y_n),~~~A_{y,n}\equiv A_y(y_n),
 \end{eqnarray}
the second term \eqref{gkt1} becomes
 \begin{eqnarray}
  -\frac{1}{4}\int_0^Ldy[(\partial_\mu A_y)^2+(\partial_yA_\mu)^2]=
  -\frac{1}{4}\sum_{n=0}^N
             \left[(\partial_\mu A_{y,n})^2
                    +\left(\frac{A_{\mu,(n+1)}-A_{\mu,n}}{a}\right)^2\right],
  \label{mass}
 \end{eqnarray}
where $a$ is a size of lattice and a periodicity, $A_{\mu,N+1}=A_{\mu,0}$, is 
assumed.

It might be instructive to show here about the gauge transformation before 
proceeding discussion. Let us consider the transformation for a bi-fundamental field as 
 \begin{eqnarray}
  \Phi_{n,n+1}\rightarrow U_n\Phi_{n,n+1}U_{n+1}^{-1}.
 \end{eqnarray}
And we write the covariant derivative as
 \begin{eqnarray}
  D_{\mu}\Phi_{n,n+1}\equiv\partial_\mu\Phi_{n,n+1}+igA_{\mu n}\Phi_{n,n+1}
                                      -ig\Phi_{n,n+1}A_{\mu,n+1}.
 \end{eqnarray}
The gauge transformations for $n$-th index are
 \begin{eqnarray}
  \Phi_{n,n+1} &\rightarrow& U_n\Phi_{n,n+1},\\
  D_{\mu}\Phi_{n,n+1} &\rightarrow&
      \partial_\mu U_n\Phi_{n,n+1}+U_n\partial_\mu\Phi_{n,n+1}
      +igA_{\mu,n}'U_n\Phi_{n,n+1}-igU_n\Phi_{n,n+1}A_{\mu,n+1}.\label{DPhi}
 \end{eqnarray}
Since the right hand side of \eqref{DPhi} should be 
$U_n(\partial_\mu\Phi_{n,n+1}+igA_{\mu,n}\Phi_{n,n+1}-ig\Phi_{n,n+1}A_{\mu,n+1})$ in the correct gauge transformations, the following relation should be 
satisfied,
 \begin{eqnarray}
  igA_{\mu n}'=U_nigA_{\mu,n}U_n^{-1}-(\partial_\mu U_n)U_n^{-1}. \label{nth}
 \end{eqnarray}
In the same manner, for the $(n+1)$-th index, we have
 \begin{eqnarray}
  &&\Phi_{n,n+1}\rightarrow\Phi_{n,n+1}U_{n+1}^{-1},\\
  &&D_\mu\Phi_{n,n+1}\rightarrow\partial_\mu\Phi_{n,n+1}U_{n+1}^{-1}
    +\Phi_{n,n+1}\partial_\mu U_{n+1}^{-1}+igA_{\mu n}\Phi_{n,n+1}U_{n+1}^{-1}
    -ig\Phi_{n,n+1}U_{n+1}^{-1}A_{\mu,n+1}'. \nonumber\\
  \label{DPhi1}
 \end{eqnarray}
Since the right hand side of \eqref{DPhi1} should be $(\partial_\mu\Phi_{n,n+1}+igA_{\mu,n}\Phi_{n,n+1}-ig\Phi_{n,n+1}A_{\mu,n+1})U_{n+1}^{-1}$, the relations
 \begin{eqnarray}
  igA_{\mu,n+1}'=U_{n+1}igA_{\mu,n+1}U_{n+1}^{-1}
                 -(\partial_\mu U_{n+1})U_{n+1}^{-1}, \label{n1th}
 \end{eqnarray}
are required. Therefore, we find from \eqref{nth} and \eqref{n1th} that both 
$n$ and $(n+1)$-th fields in a bi-fundamental one can be covariant under the 
same gauge transformation:
 \begin{eqnarray}
  D_\mu\Phi_{n,n+1}\rightarrow U_n(D_\mu\Phi_{n,n+1})U_{n+1}^{-1}.
 \end{eqnarray}
When we compactify the extra-dimensional space by $S^1$ meaning $U_N=U_0$, the 
Lagrangian of $[SU(N_c)]^N$ gauge theory in four dimensions, which is 
equivalent to $SU(N_c)$ gauge theory in five dimension, is written by
 \begin{eqnarray}
  \mathcal{L}=\sum_{n=0}^{N-1}|D_\mu\Phi_{n,n+1}|^2
                   +\mbox{tr}(\Phi_{0,1}\Phi_{1,2}\cdots\Phi_{N-1,N}).
  \label{sunn}
 \end{eqnarray}
Here, let us consider a VEV, $\langle\Phi_{n,n+1}\rangle=v\delta_{n,n+1}$, and 
expansion around the VEV as
 \begin{eqnarray}
  A=\left(
          \begin{array}{cccc}
           A_{11} & A_{12} & \cdots & A_{1N_c} \\
           \vdots & \vdots & \ddots & \vdots \\
           A_{N_c1} & A_{N_c2} & \cdots & A_{N_cN_c}
          \end{array}
         \right)
   \equiv\Phi_{n,n+1}= iv\delta_{n,n+1}+\Phi_{n,n+1}',
 \end{eqnarray} 
where the bi-fundamental field, $\Phi_{n,n+1}$, is transformed by the same $U$ 
for all $n$ as shown above, $\Phi_{n,n+1}\rightarrow U\Phi_{n,n+1}U^{-1}$. 
Under this expansion, the kinetic term for the scalar field in \eqref{sunn} can
 be rewritten by
 \begin{eqnarray}
  \sum_{n=0}^{N-1}|D_\mu\Phi_{n,n+1}|^2
  =\sum_{n=0}^{N-1}|\partial_\mu\Phi_{n,n+1}'-gv(A_{\mu,n}-A_{\mu,n+1})
                    +igA_{\mu,n}\Phi_{n,n+1}'-ig\Phi_{n,n+1}'A_{\mu,n}|^2.
  \nonumber \\
 \end{eqnarray}
Comparing this mass term of gauge field with that of \eqref{mass}, we find the 
following correspondences among the gauge coupling, VEV, size of lattice, 
compactification radius, and number of $SU(N_c)$,
 \begin{eqnarray}
  gv\leftrightarrow\frac{1}{a},~~~~~~\pi R\leftrightarrow aN. \label{corresp}
 \end{eqnarray}
Let us return to the discussion about mass term of gauge field \eqref{mass}. 
{}From the above correspondences \eqref{corresp}, it is seen that the mass term
 of gauge field can be described as
 \begin{eqnarray}
  \mathcal{L}_{\text{kin}}
  \supset\sum_{n=0}^{N-1}g^2v^2(A_{\mu,n}-A_{\mu,n+1})^2, \label{mass1}
 \end{eqnarray}
after the deconstruction. 

What about the five-dimensional Lagrangian and extra-dimensional BCs for gauge 
field under this deconstruction? We start with the following five-dimensional 
Lagrangian,
 \begin{eqnarray}
  \mathcal{L}_{5D}=-\frac{1}{2}\mbox{tr}F_{\mu\nu}F^{\mu\nu}
                          -\frac{1}{2}\mbox{tr}(\partial_\mu A_5
                          -\partial_5A_\mu+ig[A_\mu,A_5])^2.
 \end{eqnarray}
After latticizing the five-dimensional coordinate, the Lagrangian becomes
 \begin{eqnarray}
  \mathcal{L}_{5D}
  &\rightarrow& -\frac{1}{2}\sum_n^{N-1}\mbox{tr}F_{\mu\nu}^nF^{n\mu\nu}
                \nonumber\\
  &           & -\frac{1}{2}\sum_n^{N-1}\mbox{tr}
                \left|\partial_\mu A_{5,n,n+1}
                -\left(\frac{A_{\mu,n+1}-A_{\mu,n}}{a}\right)
                +ig(A_{\mu,n}A_{5,n,n+1}-A_{5,n,n+1}A_{\mu,n+1})\right|^2.
                \nonumber\\
 \end{eqnarray}
Notice that the gauge field $A_{\mu,n}$ is transformed as $A_{\mu,n}\rightarrow
 U_nA_{\mu,n}U_n^{-1}-(\partial_\mu U_n)U_n^{-1}$. It is seen that the Dirichlet and Neumann type BCs can be written by
 \begin{eqnarray}
  A_{\mu,0}=0,~~~~~~A_{\mu,N}=0, \label{D}
 \end{eqnarray}
for Dirichlet BCs at the boundaries, and 
 \begin{eqnarray}
  A_{\mu,0}=A_{\mu,1},~~~~~~A_{\mu,N-1}=A_{\mu,N}, \label{N}
 \end{eqnarray}
for Neumann ones, respectively. We comment on the scalar sector described by a 
bi-fundamental field. The general Lagrangian is given as
 \begin{eqnarray}
  \mathcal{L}=\sum_{n=1}^{N-1}\mbox{tr}|D_\mu\Phi_{n,n+1}|^2+V(x),
 \end{eqnarray}
where
 \begin{eqnarray}
  x\equiv\mbox{tr}|\Phi_{0,1}\Phi_{1,2}\cdots\Phi_{N-1,N}|^2,~~~~~~
  V(x)=\lambda(x^2-v^{2N})^2.
 \end{eqnarray}
The BCs for this bi-fundamental field are
 \begin{eqnarray}
  \Phi_{0,1}=0,~~~~~~\Phi_{N-1,N}=0,
 \end{eqnarray}
for Dirichlet BCs, and
 \begin{eqnarray}
  \Phi_{0,1}=\Phi_{1,2},~~~~~~\Phi_{N-2,N-1}=\Phi_{N-1,N},
 \end{eqnarray}
for Neumann ones, respectively. 

\begin{figure}
\begin{center}
\begin{tabular}{cccc}
\multicolumn{4}{c}{[Neumann]} \\
0 mode & 1st KK mode & 2nd KK mode & 3rd KK mode \\
\includegraphics[scale = 0.42]{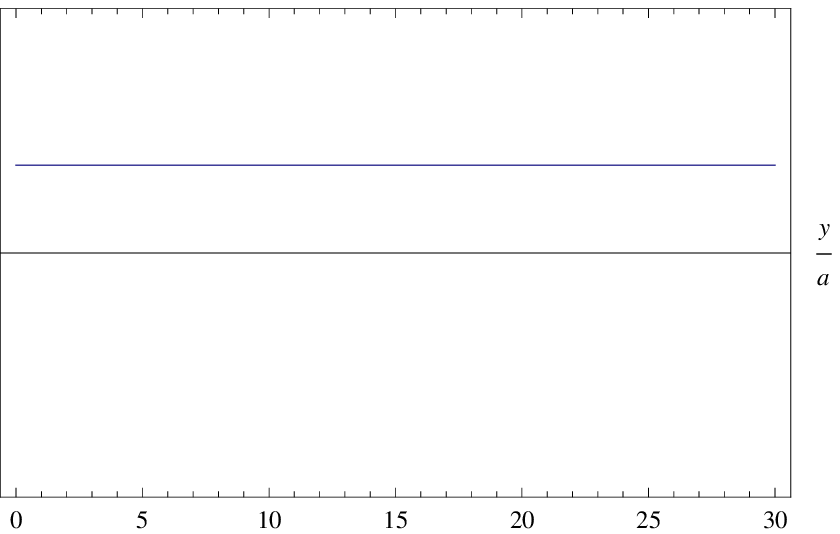} & \includegraphics[scale = 0.42]{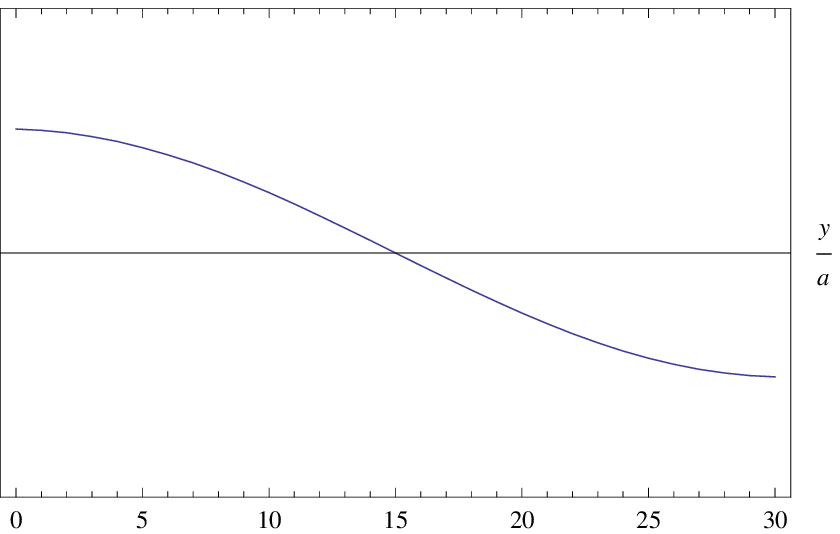} & \includegraphics[scale = 0.42]{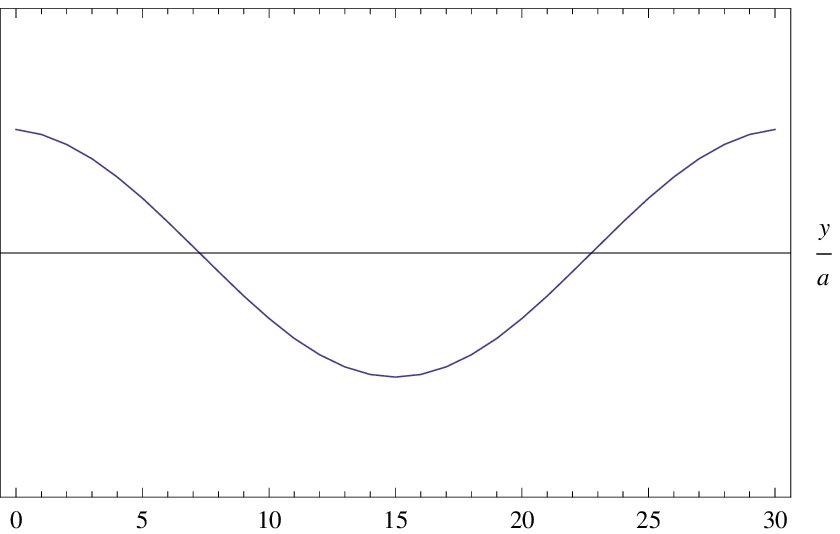} & \includegraphics[scale = 0.42]{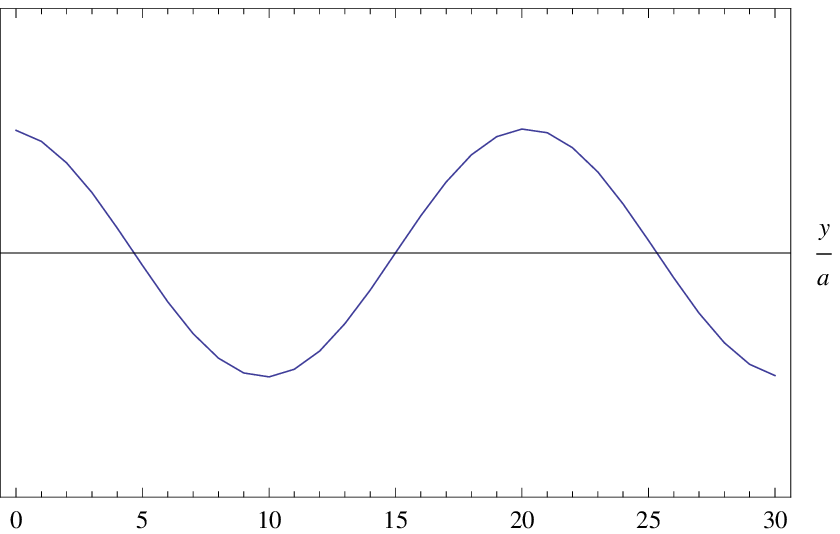} \\
\multicolumn{4}{c}{[Dirichlet]} \\
0 mode & 1st KK mode & 2nd KK mode & 3rd KK mode \\
\raisebox{1.2cm}{-} & \includegraphics[scale = 0.42]{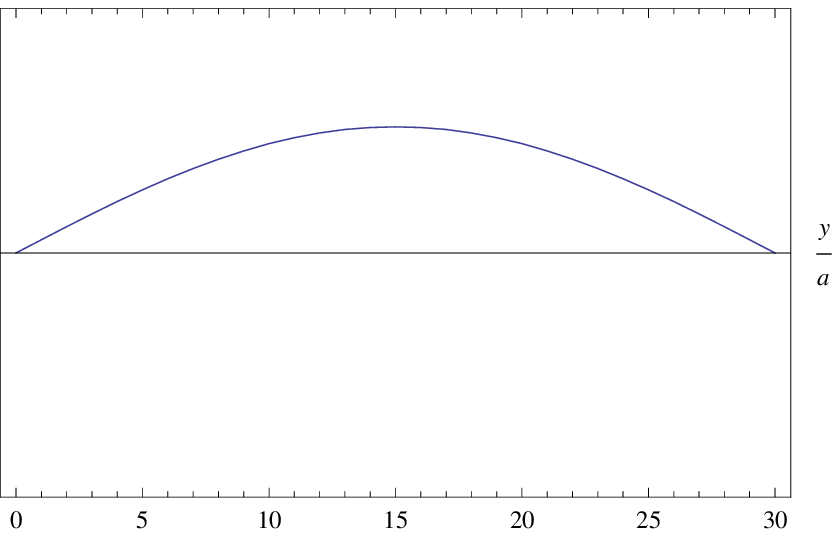} & \includegraphics[scale = 0.42]{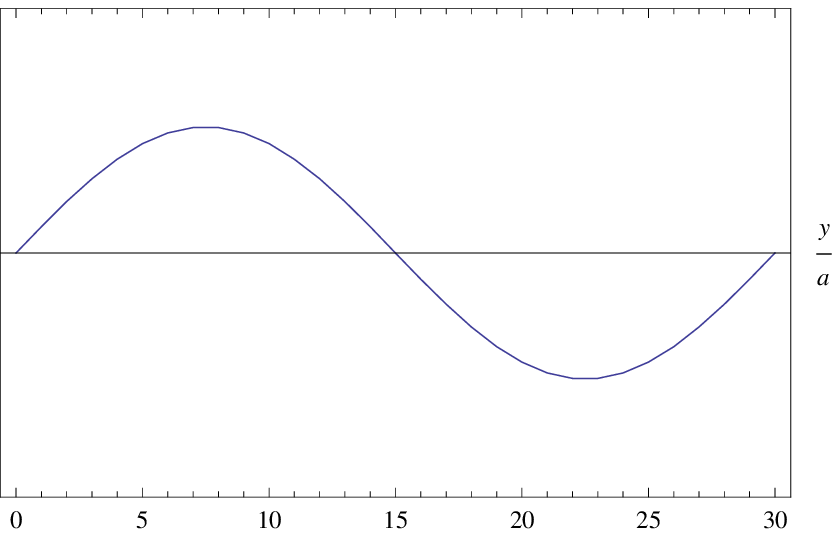} & \includegraphics[scale = 0.42]{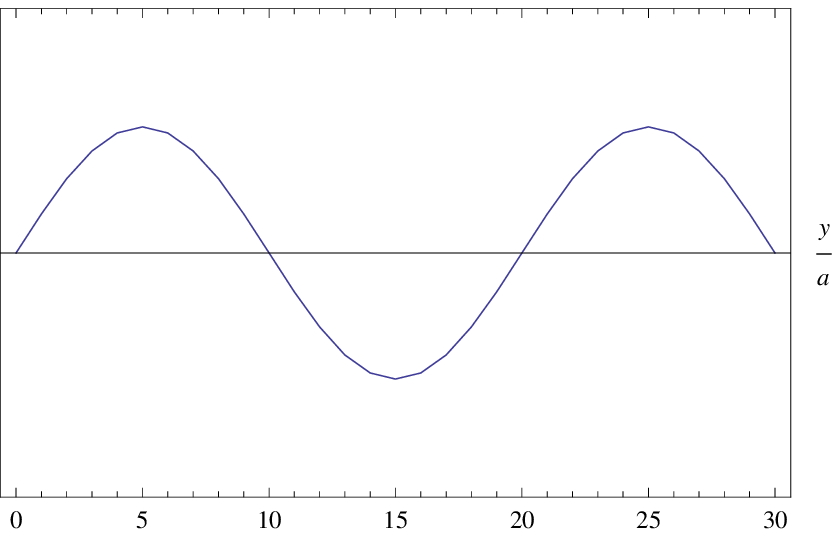} 
\end{tabular}
\end{center}
\caption{The numerical plots of the gauge field profiles for $N=30$ case in the
 deconstruction method for the flat extra-dimension. In the Dirichlet BC case, 
there is not a zero mode.}
\label{fig-B}
\end{figure}

So far, we obtained the descriptions of mass term of gauge boson \eqref{mass1} 
and BCs \eqref{D} and \eqref{N}. Finally, we show typical wave-function 
profiles of gauge field under the above each BC. Under the Neumann type BCs 
\eqref{N}, the mass term of gauge boson can be written down by
 \begin{eqnarray}
  g^2v^2(A_{\mu,1}~\cdots~A_{\mu,N})
  \left(
   \begin{array}{ccccccc}
    1      & -1     & 0      & \cdots & 0      & 0      & 0      \\
    -1     & 2      & -1     & \cdots & 0      & 0      & 0      \\
    0      & -1     & 2      & \cdots & 0      & 0      & 0      \\
    \vdots & \vdots & \vdots & \ddots & \vdots & \vdots & \vdots \\
    0      & 0      & 0      & \cdots & 2      & -1     & 0      \\
    0      & 0      & 0      & \cdots & -1     & 2      & -1     \\
    0      & 0      & 0      & \cdots & 0      & -1     & 1
   \end{array}
  \right)
  \left(
   \begin{array}{c}
    A_{\mu,1} \\
    \vdots    \\
    \vdots    \\
    \vdots    \\
    \vdots    \\
    A_{\mu,N}
   \end{array}
  \right),\label{deco-neumann}
 \end{eqnarray}
where we consider the sum up to $N$. For the Dirichlet BC, the mass term is 
described as
 \begin{eqnarray}
  g^2v^2(A_{\mu,1}~\cdots~A_{\mu,N})
  \left(
   \begin{array}{ccccccc}
    2      & -1     & 0      & \cdots & 0      & 0      & 0      \\
    -1     & 2      & -1     & \cdots & 0      & 0      & 0      \\
    0      & -1     & 2      & \cdots & 0      & 0      & 0      \\
    \vdots & \vdots & \vdots & \ddots & \vdots & \vdots & \vdots \\
    0      & 0      & 0      & \cdots & 2      & -1     & 0      \\
    0      & 0      & 0      & \cdots & -1     & 2      & -1     \\
    0      & 0      & 0      & \cdots & 0      & -1     & 2
   \end{array}
  \right)
  \left(
   \begin{array}{c}
    A_{\mu,1} \\
    \vdots    \\
    \vdots    \\
    \vdots    \\
    \vdots    \\
    A_{\mu,N}
   \end{array}
  \right).\label{deco-dirichlet}
 \end{eqnarray}
In the deconstruction method, one can obtain gauge field profiles in an 
extra-dimensional direction by taking the wave-function in the basis of mass 
eigenstate. The basis of eigenstate are changed by operating an unitary matrix,
 $U_A$, which diagonalizes the mass matrix given in \eqref{deco-neumann} or 
\eqref{deco-dirichlet}, as
 \begin{eqnarray}
  g^2v^2(A_{\mu,1}~\cdots~A_{\mu,N})U_A^\dagger
   \left(
    \begin{array}{cccc}
     m_{A^{(0)}}^2 & 0             & \cdots & 0               \\
     0             & m_{A^{(1)}}^2 & \cdots & 0               \\
     \vdots        & \vdots        & \ddots & \vdots          \\
     0             & 0             & \cdots & m_{A^{(N-1)}}^2
    \end{array}
   \right)U_A
   \left(
    \begin{array}{c}
     A_{\mu,1} \\
     \vdots \\
     A_{\mu,N}
     \end{array}
    \right).
 \end{eqnarray}
Therefore, the mass eigenstate for each KK mode can be described by
 \begin{eqnarray}
  \left(
   \begin{array}{c}
    A_{\mu}^{(0)}   \\
    A_{\mu}^{(1)}   \\
    \vdots          \\
    A_{\mu}^{(N-1)}
   \end{array}
  \right)&=&U_A
  \left(
   \begin{array}{c}
    A_{\mu,1} \\
    A_{\mu,2} \\
    \vdots    \\
    A_{\mu,N}
   \end{array}
  \right) \\
  &=&\left(
        \begin{array}{c}
         (U_A)_{11}A_\mu(a)+(U_A)_{12}A_\mu(2a)+\cdots+(U_A)_{1N}A_\mu(Na) \\
         (U_A)_{21}A_\mu(a)+(U_A)_{22}A_\mu(2a)+\cdots+(U_A)_{2N}A_\mu(Na) \\
         \vdots                                                \\
         (U_A)_{N1}A_\mu(a)+(U_A)_{N2}A_\mu(2a)+\cdots+(U_A)_{NN}A_\mu(Na)
   \end{array}
  \right).
 \end{eqnarray}
Since this description is one after the dimensional reduction of 
extra-dimension, the wave-function profile for the field $A_\mu^{(n)}$ is 
composed of $(a,U_{n1}),$ $(2a,U_{n2}),\cdots,$ $(Na,U_{NN})$. The numerical 
plots of the wave-function profile for $N=30$ case are shown in Fig. 
\ref{fig-B}.

\begin{figure}
\begin{center}
\begin{tabular}{cccc}
\multicolumn{4}{c}{[Neumann]} \\
0 mode & 1st KK mode & 2nd KK mode & 3rd KK mode \\
\includegraphics[scale = 0.42]{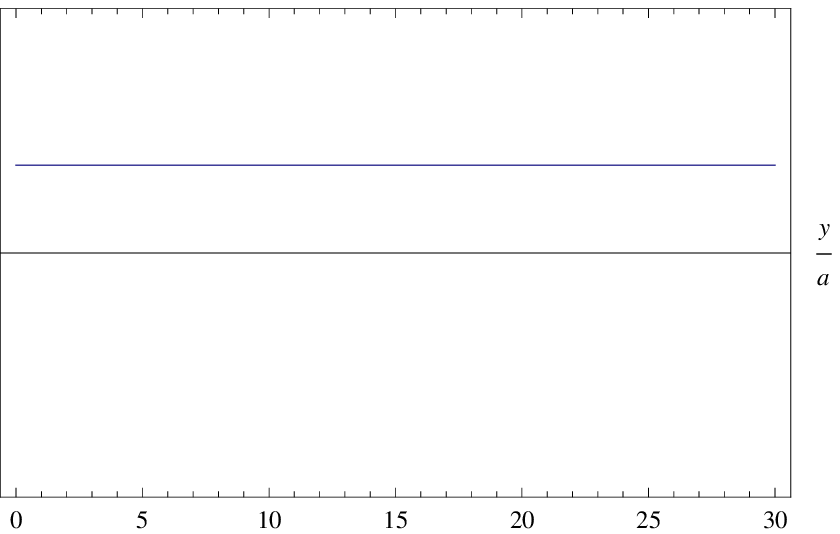} & \includegraphics[scale = 0.42]{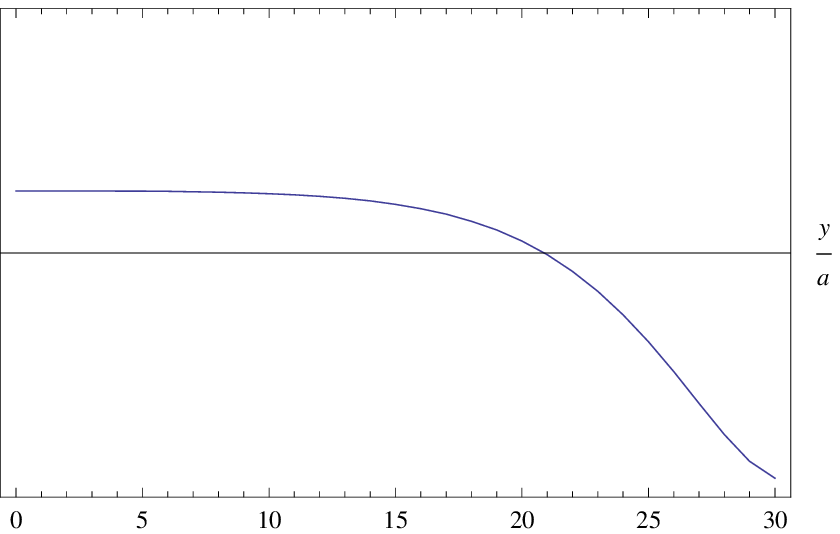} & \includegraphics[scale = 0.42]{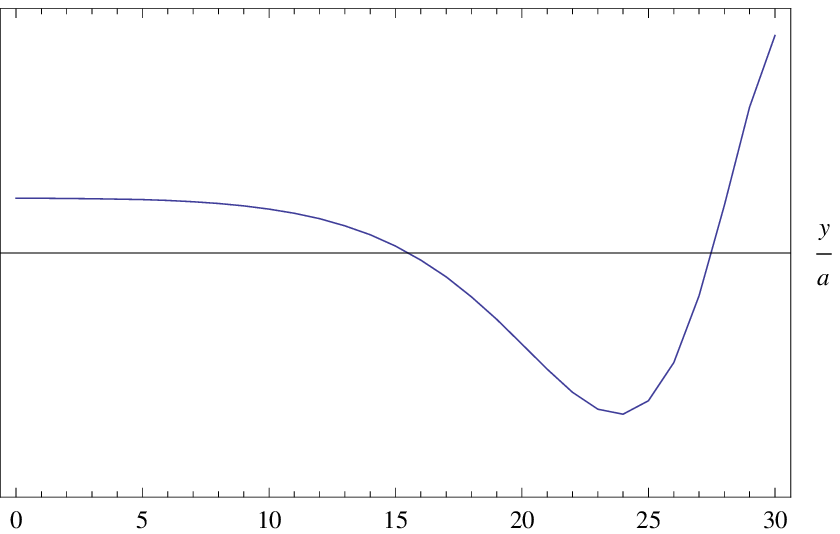} & \includegraphics[scale = 0.42]{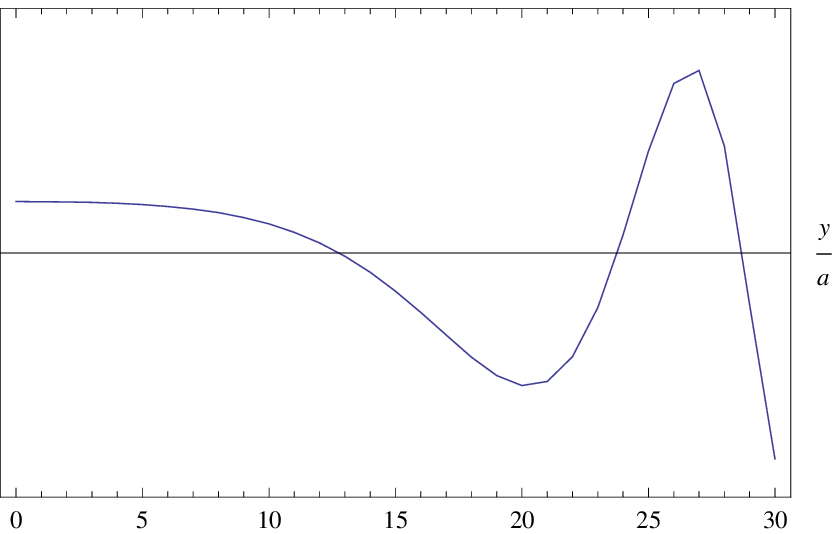} \\
\multicolumn{4}{c}{[Dirichlet]} \\
0 mode & 1st KK mode & 2nd KK mode & 3rd KK mode \\
\raisebox{1.2cm}{-} & \includegraphics[scale = 0.42]{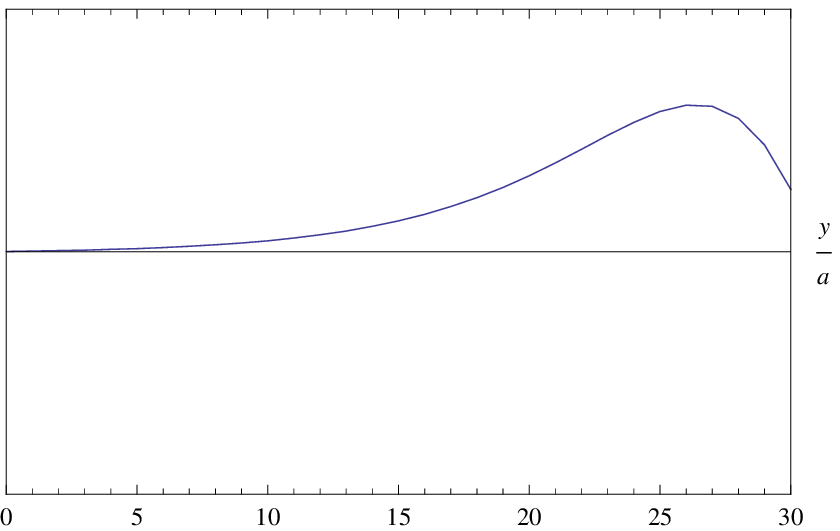} & \includegraphics[scale = 0.42]{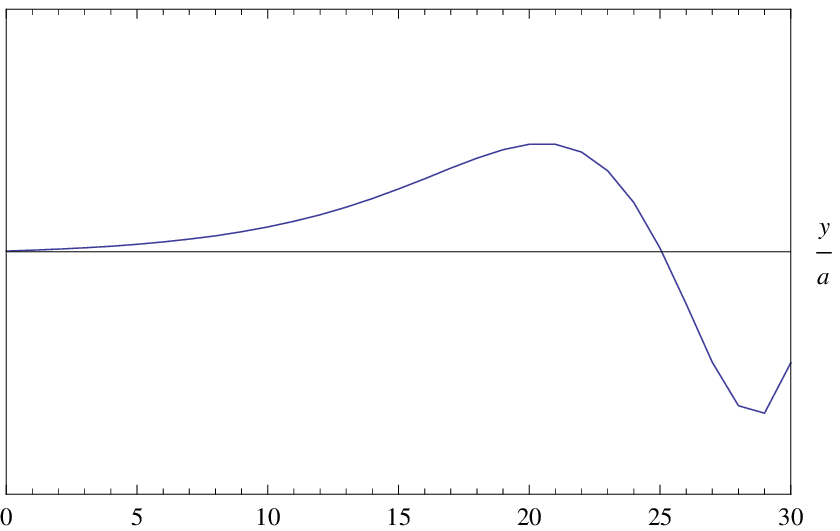} & \includegraphics[scale = 0.42]{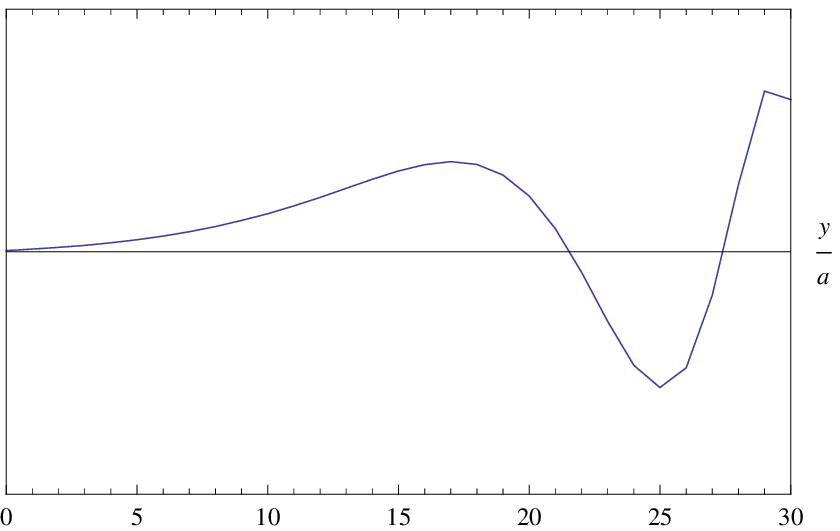} 
\end{tabular}
\end{center}
\caption{The numerical plots of the gauge field profiles for $N=30$ case in the
 deconstruction method for the warped extra-dimension. In the Dirichlet BC 
case, there is not a zero mode.}
\label{fig-C}
\end{figure}

Finally, we present an example in the warped case\footnote{See ref. 
\cite{deBlas:2006fz} for deconstructing gauge theories in AdS$_5$.}. The mass 
matrices of gauge boson in the Neumann \eqref{deco-neumann} and Dirichlet 
\eqref{deco-dirichlet} are modified to
 \begin{eqnarray}
  \left(
   \begin{array}{ccccc}
    e^{-2ka}      & -e^{-2ka} & \cdots & 0      & 0      \\
    -e^{-2ka}     & e^{-3ka}F(ka) & \cdots & 0      & 0      \\
    \vdots & \vdots & \ddots & \vdots & \vdots \\
    0      & 0 & \cdots & e^{-(2(N-1)-1)ka}F(ka)      & -e^{-2(N-1)ka}     \\
    0      & 0 & \cdots & -e^{-2(N-1)ka}    & e^{-2(N-1)ka}
   \end{array}
  \right),
 \end{eqnarray}
and
 \begin{eqnarray}
  \left(
   \begin{array}{ccccc}
    e^{-ka}F(ka) & -e^{-2ka} & \cdots & 0      
& 0      \\
    -e^{-2ka}     & e^{-3ka}F(ka) & \cdots & 0      
& 0      \\
    \vdots & \vdots & \ddots & \vdots 
& \vdots \\
    0      & 0 & \cdots & e^{-(2(N-1)-1)ka}F(ka) & -e^{-2(N-1)ka}     \\
    0      & 0 & \cdots & -e^{-2(N-1)ka}    & e^{-2(N-1)ka}F(ka)
   \end{array}
  \right),
 \end{eqnarray}
respectively, where $F(ka)\equiv e^{-ka}+e^{ka}$. The numerical plots of the 
wave-function profile for $N=30$ case are shown in Fig. \ref{fig-C}.


\end{document}